\documentclass[fleqn,usenatbib]{mnras}

\usepackage{newtxtext,newtxmath}

\usepackage[T1]{fontenc}
\usepackage{lmodern}

\usepackage{graphicx}	
\usepackage{amsmath}	
\usepackage{amssymb}	
\usepackage{booktabs}
\usepackage{cleveref}

\crefname{equation}{Eq.}{Eqs.}
\Crefname{equation}{Equation}{Equations}
\crefname{figure}{Fig.}{Figs.}
\Crefname{figure}{Figure}{Figures}
\crefname{table}{Table}{Tables}
\Crefname{table}{Table}{Tables}
\crefname{section}{Section}{Sections}
\Crefname{section}{Section}{Sections}

\title[Optical EVPA rotations in blazars: testing a stochastic variability 
       model with RoboPol data]
      {Optical EVPA rotations in blazars: testing a stochastic variability 
       model with RoboPol data}

\author[S. Kiehlmann et al.]{
S.~Kiehlmann,$^{1,2,3}$\thanks{E-mail: skiehl@caltech.edu}
D.~Blinov,$^{4,5}$
T.J.~Pearson,$^{3}$
I.~Liodakis$^{4,6,7}$
\\
$^{1}$Aalto University Mets\"ahovi Radio Observatory, Mets\"ahovintie 114, 02540 Kylm\"al\"a, Finland\\
$^{2}$Aalto University Department of Radio Science and Engineering, PL 13000, 00076 Aalto, Finland\\
$^{3}$Owens Valley Radio Observatory, California Institute of Technology, Pasadena, CA 91125, USA\\
$^{4}$Department of Physics and Institute for Theoretical and Computational Physics, University of Crete, 71003, Heraklion, Greece\\
$^{5}$Astronomical Institute, St. Petersburg State University, Universitetsky pr. 28, Petrodvoretz, 198504 St. Petersburg, Russia\\
$^{6}$Foundation for Research and Technology - Hellas, IESL, Voutes, GR-71110 Heraklion, Greece\\
$^{7}$KIPAC, Stanford University, 452 Lomita Mall, Stanford, CA 94305, USA\\
}

\date{Accepted 2017 August 17. Received 2017 August 16; in original form 2017 June 8.}

\pubyear{2017}

\begin{document}
\label{firstpage}
\pagerange{\pageref{firstpage}--\pageref{lastpage}}
\maketitle

\begin{abstract}
  We identify rotations of the polarization angle in a sample of blazars 
  observed for three seasons with the RoboPol instrument.
  A simplistic stochastic variability model is tested against this sample of 
  rotation events.
  The model is capable of producing samples of rotations with parameters 
  similar to the observed ones, but fails to reproduce the polarization 
  fraction at the same time. 
  Even though we can neither accept nor conclusively reject the model, we point
  out various aspects of the observations that are fully consistent with a
  random walk process.
\end{abstract}


\section{Introduction}
\label{sec:intro}

The variability of the optical polarization fraction and electric vector 
position angle (EVPA) in blazars has been known since the earliest observations
\citep{1967ApJ...148L..53K}.
Large, continuous rotations of the EVPA have been reported in various objects,
e.g., OJ~287 \citep{1988A&A...190L...8K}, BL~Lac \citep{2008Natur.452..966M},
PKS~$1510{-}089$ \citep{2010ApJ...710L.126M,2014A&A...569A..46A}, and 3C~279
\citep[e.g.][]{2008A&A...492..389L,2014A&A...567A..41A,2016AA26A...590A..10K}.
Some rotations of the optical EVPA have been reported to be associated with 
gamma-ray flares 
\citep[e.g.][]{2008Natur.452..966M,2013ApJ...768...40L,2014A&A...567A..41A}, 
suggesting a potential physical connection between these events.

Various mechanisms have been proposed to explain EVPA rotations.
Deterministic models typically assume an ordered magnetic field structure.
Examples of such models include the superposition of two or more emission 
regions \citep{1984MNRAS.211..497H}, a bend of the jet 
\citep{2010IJMPD..19..701N}, and the progression of an emission feature on a 
helical streamline \citep{2008Natur.452..966M} or a shock 
\citep{1985ApJ...289..188K,2014ApJ...789...66Z,2015ApJ...804...58Z} passing 
through a helical magnetic field.
Stochastic models on the other hand assume a tangled magnetic field structure 
and model a turbulent flow in terms of random cells
\citep[e.g.][]{1985ApJ...290..627J,2014ApJ...780...87M,2016AA26A...590A..10K}.
The most recent models of \citet{2014ApJ...789...66Z} and 
\citet{2014ApJ...780...87M} focus particularly on the broad-band variability 
of the polarized and total flux and the potential connection between optical 
EVPA rotations and high-energy flaring activity.

The RoboPol\footnote{\url{http://robopol.org/}} project, an optical 
polarization monitoring campaign of an unbiased sample of blazars 
\citep{2014MNRAS.442.1693P}, was initiated to study the optical EVPA rotation 
phenomenon in blazars on a statistical basis and in particular to address the 
question, whether these rotations are physically connected to the gamma-ray 
flaring process.
\citet{2015MNRAS.453.1669B} showed that some but not all rotations in the 
first season of the RoboPol data could be produced by a random walk process.
They also argued that some rotations occur contemporaneously with gamma-ray 
flares and that the contemporaneity is unlikely to be happening by chance, 
implying a physical connection between the two events.
In comparison, based on a single case study of 3C~279,
\citet{2016AA26A...590A..10K} demonstrated that even within the same object 
some EVPA rotations may be explained by a random walk process, while others 
cannot.
Based on the first and second season data of the RoboPol project,
\citet{2016MNRAS.457.2252B} showed that the average polarization fraction
during rotations is lower in comparison to non-rotation periods and that it is
correlated with the EVPA rotation rate.
Additionally, they demonstrated that rotation amplitudes and durations have 
upper limits.
Based on three seasons of data,
\citet{2016MNRAS.462.1775B} claim that only a fraction of
blazars exhibit large rotations of the EVPA frequently, while others rarely or 
never show such events.
\citet{2016MNRAS.463.3365A} have demonstrated that the EVPA is on average more 
variable in low synchrotron peaked (LSP) sources than in high synchrotron 
peaked (HSP) sources.

In this study we use the full data set of three seasons of RoboPol blazar 
monitoring to test a random walk process scenario.
In contrast to the case-by-case study of EVPA rotations in 3C~279 by 
\citet{2016AA26A...590A..10K} here we follow a statistical approach of testing 
an entire sample of blazars against a random walk process.
With three seasons of data, which have significantly increased the 
number of observed EVPA rotations, the RoboPol data provide a solid basis for 
a statistical study.
In particular we test whether several of the relations found in 
\citet{2016MNRAS.457.2252B} and \citet{2016MNRAS.462.1775B}
can be explained by a random walk process.


\section{Data and data processing}
\label{sec:data}

The RoboPol main sample is an unbiased, gamma-ray photon-flux-limited sample 
of 62~gamma-ray-loud blazars, selected from the \emph{Fermi}-LAT Second Source
Catalogue \citep[2FGL,][]{2012ApJS..199...31N}.
The sample selection is discussed in detail in \citet{2014MNRAS.442.1693P} and
we follow their source nomenclature.
These objects were continuously monitored with the RoboPol polarimeter at the 
1.3~meter telescope of the Skinakas observatory for three seasons: May~2013 --
November~2013, April~2014 -- November~2014, and May~2015 -- November~2015.
\citet{2014MNRAS.442.1706K} describe features of the instrument and the 
pipeline used for the data reduction.
Detailed information on the observations for each season is given in
\citet{2015MNRAS.453.1669B,2016MNRAS.457.2252B}, and
\citet{2016MNRAS.462.1775B}.
Additional quality checks, applied to the data after the automated data 
reduction, are described in \citet{2016MNRAS.463.3365A}.

19~sources in the sample were occasionally observed two to four times during a
single night.
In 13~of those sources we do not observe significant variability in both the 
EVPA and at least one Stokes parameter at these time-scales.
Here, we average the fractional Stokes parameters $q = Q/I$, $u = U/I$
within time intervals of half a day without removing significant variability.
Otherwise some measurements of the rotation variation estimator (explained in 
\cref{sec:rotations}) would be strongly affected by extreme rotation rates
induced by observational noise on very short time-scales.
Six sources with intra-night observations occasionally show significant
variability in the EVPA and at least one Stokes parameter: 
RBPL\,J1555+1111, RBPL\,J1653+3945, RBPL\,J2202+4216, RBPL\,J2232+1143,
RBPL\,J2243+2021, RBPL\,J2253+1608.
Also for these sources we average fractional Stokes parameters within time
intervals of half a day, thus averaging out real intrinsic variability.
This is done to avoid complications with the model (\cref{sec:rwsim} and
following), which cannot cover arbitrarily short time-scales due to its 
discrete set up of cells and variability.
None of the results and conclusions based on the model comparison are affected
by the intra-night averaging.
A publication that addresses how the identification of rotations and their
characteristic parameters depend on the time sampling is in preparation.

We select a duration of 50~days as an upper limit for the allowed time step
between data points and split the data gaps larger than 50~days.
In the following we refer to these subsets of data as \emph{observing periods}.
For each object, we treat each period as an individual data set when 
searching for rotations of the polarization angle.
Since we are focusing on a study of variability, we include only periods that
contain at least four data points.
Following these criteria, each object is observed for at least one period, up
to four periods (cf.~\cref{tab:obsperiods}), with a total sum of 183~periods.
The durations of the observing periods range from 12--197~days
(\cref{fig:distmodelpar}, panel~a), consisting of 4--54 data points.
The time steps during the selected periods range from 0.8--50~days
(\cref{fig:distmodelpar}, panel~b).

\begin{table}
  \caption{How many objects of the RoboPol main sample have been observed 
    for how many observing periods during the three seasons of observations.}
  \label{tab:obsperiods}
  \begin{tabular*}{\columnwidth}{@{\extracolsep{\fill} } l l}
    \toprule
    Number of objects    & Number of obs. periods  \\
    \midrule
     1  & 1  \\
     2  & 2  \\
    58  & 3  \\
     1  & 4  \\
    \bottomrule
  \end{tabular*}
\end{table}

The electric vector position angle (EVPA), $\chi$, is derived from the 
fractional Stokes parameters $q$ and $u$ \citep[][Eq.~6]{2014MNRAS.442.1706K}.
The measured EVPA is limited to an interval of $180\,\mathrm{deg}$.
Therefore, differences between EVPA data points are measured modulo 
$180\,\mathrm{deg}$ and are, thus, ambiguous.
Under the assumption of minimal variation, we shift the measured EVPA data 
points by multiples of $180\,\mathrm{deg}$ such that the difference between 
adjacent data points is always smaller than $90\,\mathrm{deg}$.
The following data analysis is entirely based on this shifted EVPA curve.
\citet{2015MNRAS.453.1669B,2016MNRAS.457.2252B,2016MNRAS.462.1775B} used a
different method that shifted data points only if the variability was 
significant.
\cite{2016AA26A...590A..10K} have shown that taking the uncertainties into 
consideration yields inconsistent results for the adjusted EVPA curve.
We point out that the two different methods yield different EVPA curves only 
for the two sources RBPL\,J0841+7053 and RBPL\,J1058+5628.
The rotations reported in
\citet{2015MNRAS.453.1669B,2016MNRAS.457.2252B,2016MNRAS.462.1775B} are not 
affected by the choice of method.

\section{Data analysis}
\label{sec:dataanalysis}

In this section, we define our quantitative criteria for a rotation of the 
EVPA.
Then rotations in the RoboPol main sample are identified and characterized
according to various parameters.
Additionally, we investigate the behaviour of the polarization fraction during
rotation and non-rotation periods.

\subsection{Definition of an EVPA rotation}
\label{sec:rotationdefinition}

We use the method introduced by \cite{2016AA26A...590A..10K} to identify 
periods of continuous EVPA rotations in the data and simulations consistently.
This method identifies periods during which the EVPA changes continuously in
one direction.
Counter-rotations at the level of the uncertainties do not interrupt a 
rotation.
Only significant variability with a change of direction ends a detected 
rotation.
A rotation between two data points $\chi_i$, $\chi_j$ with uncertainties
$\sigma_{\chi, i}$, $\sigma_{\chi, j}$ is considered significant when
\begin{align}
  \left| \chi_i - \chi_{j} \right| > \varsigma \sqrt{\sigma_{\chi, i}^2 + \sigma_{\chi, j}^2}.
\end{align}
The factor $\varsigma$ scales to what extent variability is interpreted as 
intrinsic or noise induced.
We choose $\varsigma = 1$.

As we employ a method different from the one used in 
\citet{2015MNRAS.453.1669B,2016MNRAS.457.2252B,2016MNRAS.462.1775B}
the rotations discussed here may differ from the ones previously reported for 
the RoboPol main sample.
The method used here is likely to identify more rotations, as it decreases the 
chance of false non-detections when observational noise apparently changes the 
direction of the intrinsic variability.
On the other hand, this method comes at the cost of increasing the chance of 
false positive detections.
This could be the case when we observe intrinsic variability that is below the
uncertainties.
This intrinsic variability would break our definition of uni-directional EVPA
variability, if it is a counter-rotation.
But as it is insignificant, this variability is interpreted as noise and not
as intrinsic and will not end a rotation.
The advantages and disadvantages of both methods and consequences of the 
method selection will be discussed in more detail in a future paper.
For this study it is merely important to use one method consistently.

\begin{figure*}
  \centering
  \includegraphics[width=\textwidth]{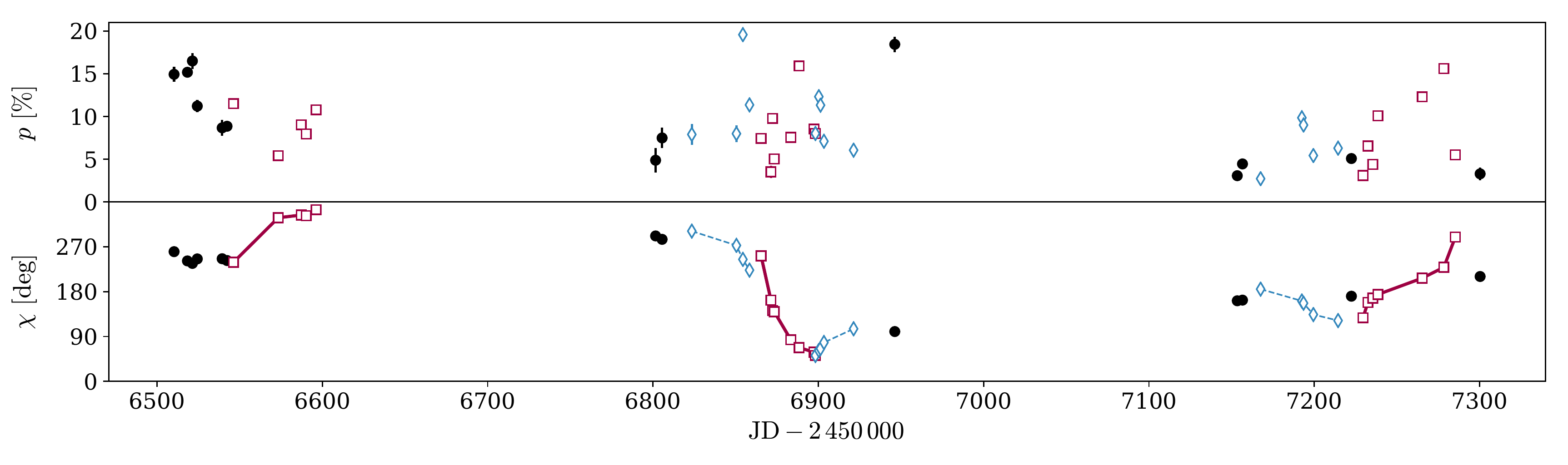}
  \caption{
    Polarization fraction and EVPA of RBPL\,J1800+7828. Coloured data points, 
    connected by lines mark periods of small ($\Delta\chi < 90\,\mathrm{deg}$,
    blue diamonds, dashed lines) and large ($\Delta\chi \geq 90\,\mathrm{deg}$, 
    red squares, solid lines) rotations. Black points are not part of
    a rotation period.}
  \label{fig:rotationexample}
\end{figure*}

Rotations are identified after adjusting the EVPA curve for the
$180\,\mathrm{deg}$~ambiguity.
We identify rotations for each observing period individually, i.e. we do not
consider rotation to continue after a gap of $\geq 50$~days.
As in \citet{2015MNRAS.453.1669B,2016MNRAS.457.2252B,2016MNRAS.462.1775B}, 
we accept only rotations that contain at least four data points.
\citet{2015MNRAS.453.1669B,2016MNRAS.457.2252B,2016MNRAS.462.1775B} furthermore 
considered only rotations with large amplitudes
$\Delta\chi \geq 90\,\mathrm{deg}$.
We do not apply a cut on the amplitude in general.
In the following we discuss results for two schemes of identifying rotations: 
first, accepting all rotations with arbitrary amplitudes and,
second, accepting only rotations with large amplitudes
($\Delta\chi \geq 90\,\mathrm{deg}$).
We designate the former as \emph{rotations} and the latter as \emph{large
rotations}.
Rotations with amplitudes $\Delta\chi < 90\,\mathrm{deg}$ are referred to as 
\emph{small rotations}.
\Cref{fig:rotationexample} shows an example of identified rotation periods
for RoboPol object RBPL\,J1800+7828.
Red data points mark periods of large rotations, blue points mark small 
rotations, and black points are not identified as part of a rotation.

\begin{figure*}
  \centering
  \includegraphics[width=\textwidth]{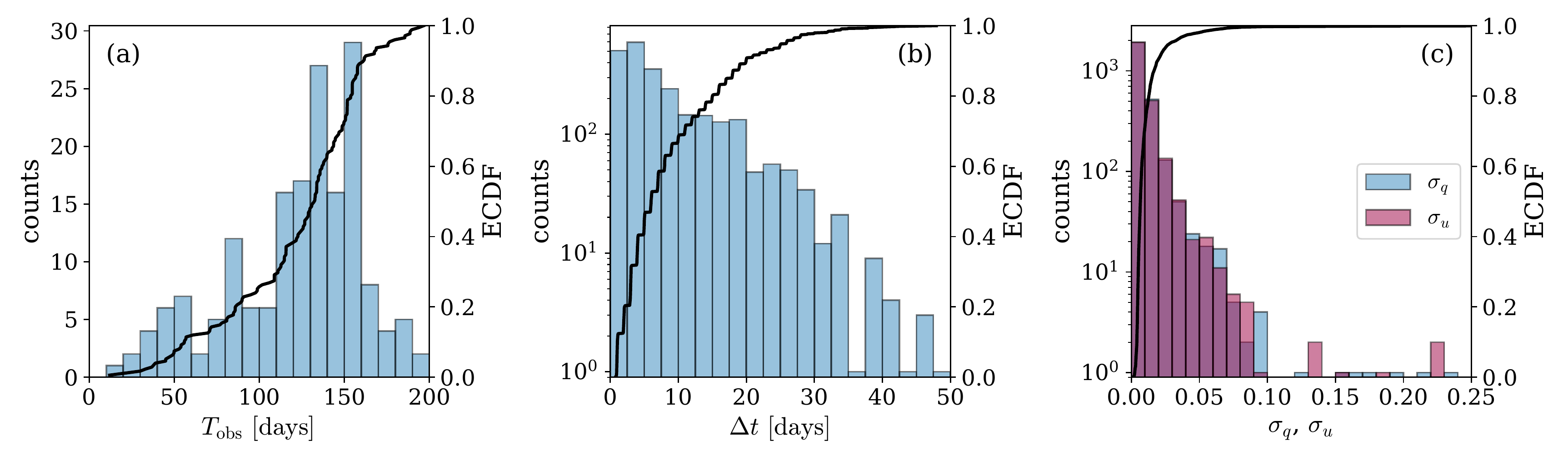}
  \caption{
    Empirical distributions based on the RoboPol main sample used for the 
    random simulations.
    \emph{Panel~a:} Duration of observing periods.
    \emph{Panel~b:} Time steps during observing periods.
    \emph{Panel~c:} Uncertainties of the fractional Stokes parameters $q$ and 
    $u$ (blue and red histogram). The  empirical cumulative distribution 
    function (ECDF, black line) is based on the combined uncertainties of $q$ 
    and $u$.} 
  \label{fig:distmodelpar}
\end{figure*}

The identification of rotations depends on the sampling and the measurement
uncertainties.
There is always a chance of identifying rotations wrongly due to the 
uncertainties or to miss them due to uncertainties or insufficient time 
sampling.
Therefore, in the comparison of data and models the time sampling and 
measurement uncertainties, as shown in \cref{fig:distmodelpar}, have to be 
accounted for in the model simulations.

\subsection{Characteristics of the rotations}
\label{sec:rotations}

Following the criteria above, we have identified 343~rotations in the three 
seasons of the RoboPol main sample observations, 98~of which are large 
rotations (\cref{tab:rotnum}).
The third column in \cref{tab:rotnum} shows the frequency of rotations per 
$100$~days, averaged over the entire sample.
This estimate is based on the sum of the observing periods 
$T_\mathrm{obs,tot} = 21783.7$~days.
Durations of the observing periods are not taken into account.
Therefore, the estimated frequencies depend not only on the intrinsic behaviour
of the sources, but also on the observing schedule.

\begin{table}
  \caption{
    Absolute number and estimated frequency of rotations observed in the 
    RoboPol main sample and number of sources showing rotations with certain 
    amplitudes.}
  \label{tab:rotnum}
  \begin{tabular*}{\columnwidth}{@{\extracolsep{\fill} } r r r r r }
    \toprule
    Amplitudes    & \multicolumn{2}{c}{Number of rotations} & \multicolumn{2}{c}{Number of sources}  \\
                  & absolute   & per $100$~days             & absolute  & relative    \\
    \midrule
    all                       & 343  &  1.57  & 61  & 98\%  \\
    $\geq  90\,\mathrm{deg}$  &  98  &  0.45  & 43  & 69\%  \\
    $\geq 180\,\mathrm{deg}$  &  25  &  0.11  & 15  & 24\%  \\
    $\geq 270\,\mathrm{deg}$  &   3  &  0.01  &  3  &  5\%  \\
    $\geq 360\,\mathrm{deg}$  &   1  & <0.01  &  1  &  2\%  \\
    \bottomrule
  \end{tabular*}
\end{table}

We measure four parameters that characterize the rotations.
The \emph{amplitude} of a rotation is
$\Delta\chi = \max(\chi_i) - \min(\chi_i)$, where $\chi_i$ are the data points
of an identified rotation.
The \emph{duration} $T_\mathrm{rot}$ is the time passing between the start and 
end of the rotation.
We estimate the average \emph{rotation rate} through 
$\Delta\chi / T_\mathrm{rot}$.
And we use the \emph{variation estimator} $s_\chi$ defined in 
\citet{2016AA26A...590A..10K} to quantify the smoothness of the EVPA curve:
\begin{align}
  s_\chi = \left< \left| s_i \right| \right> &= \left< \left| \left( 
  \frac{\Delta \chi}{\Delta t} \right)_i - \left< \left( 
  \frac{\Delta \chi}{\Delta t} \right) \right> \right| \right> 
  \label{eq:varest} \\
  \text{with\ } \left( \frac{\Delta \chi}{\Delta t} \right)_i &= 
  \frac{\chi_i - \chi_{i-1}}{t_i - t_{i-1}},
\end{align}
where $\left< \cdot \right>$ denotes the mean over all $i$.
A larger value reflects stronger variability and a less smooth curve.

\subsubsection{Limits and biases in the rotation parameters}
\label{sec:rotparbias}

The variation estimator is affected by observational errors and intrinsic 
curvature of the EVPA curve.
Both biases increase $s_\chi$ \citep{2016AA26A...590A..10K}.
The rotation amplitude, duration and rate are affected by the time sampling and 
limited observing periods.
The measured rotation amplitude and duration are generally lower limits.
A detected rotation probably starts between the first data point of the 
measured rotation and the preceding data point and continues between the last 
and the following data point.
Accordingly, the estimates of the average rotation rate are affected and the 
actual values could be larger or smaller.
The uncertainties of these measurements depend on the time sampling and the
source intrinsic variability.

Most critical are cases in which the detected rotation starts or ends with
the observing period.
The actual rotation could extend far beyond this period.
The limited duration of the observing periods (\cref{fig:distmodelpar}, 
panel~a) puts a strict constraint on the measurable duration of rotations and 
the amplitudes and durations may as a result be underestimated and rotation
rates wrongly estimated accordingly.
The majority of rotations ($72\%$) is fully covered by the corresponding
observing periods.
The measured parameters of these rotations are solely limited by the limited
time sampling.
$26\%$ of the rotations either start or end with the observing period and eight
identified rotations cover the entire period, during which they are observed.

The estimated rotation parameters are generally limited or biased and depend
not only on the intrinsic variability of the sources but also on the scheduling
of observations and the measurement uncertainties.
In the comparison of the data and simulations these limitations have to be 
accounted for by reproducing the time sampling, observing periods, and
uncertainties of the data in the simulations.

\subsubsection{Distributions of the rotation parameter}
\label{sec:rotpardist}

\begin{table}
  \caption{
    Minimum, maximum, mean, and median values of the rotation amplitudes
    (\emph{col.~2)}), durations (\emph{col.~3)}), rates (\emph{col.~4)}), and 
    variation estimators (\emph{col.~5)}) observed in the RoboPol main sample.}
  \label{tab:rotpar}
  \begin{tabular*}{\columnwidth}{@{\extracolsep{\fill} } l r r r r}
    \toprule
    & $\Delta\chi$      & $T_\mathrm{rot}$   & $\Delta\chi / T_\mathrm{rot}$  & $s_\chi$  \\
    & $[\mathrm{deg}]$  & $[\mathrm{days}]$  & $[\mathrm{deg}/\mathrm{day}]$  & $[\mathrm{deg}/\mathrm{day}]$ \\
    \midrule
    Min.:    &   $4$  &   $3$  &  $0.1$  &  $0.1$\\
    Max.:    & $423$  & $120$  & $52.1$  & $26.8$\\
    Mean:    &  $70$  &  $33$  &  $3.1$  &  $3.3$\\
    Median:  &  $49$  &  $29$  &  $1.9$  &  $1.9$\\
    \bottomrule
  \end{tabular*}
\end{table}

\begin{figure}
  \centering
  \includegraphics[width=\columnwidth]{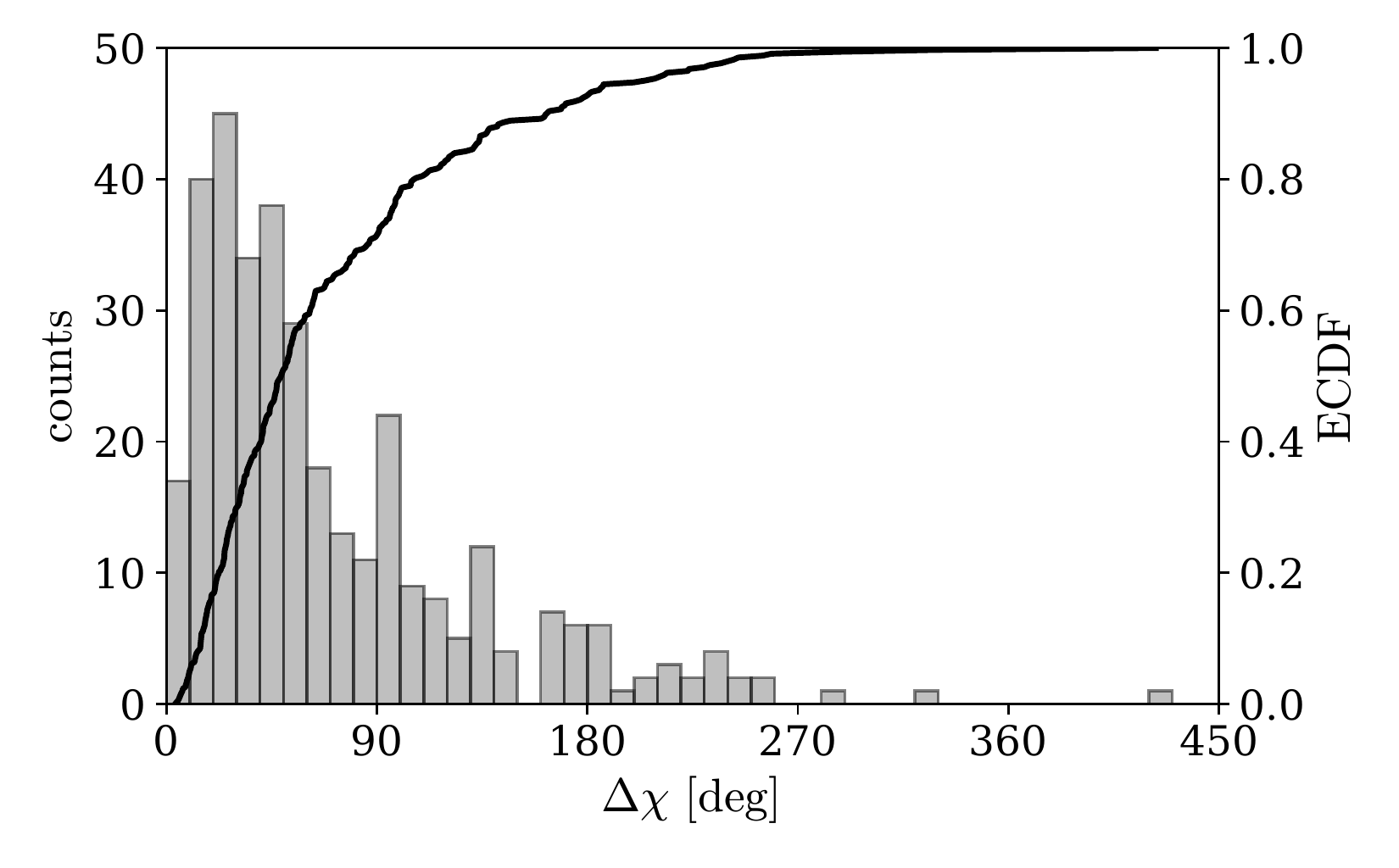}
  \caption{Distribution of rotation amplitudes in the RoboPol main sample.}
  \label{fig:distampl}
\end{figure}

\begin{figure}
  \centering
  \includegraphics[width=\columnwidth]{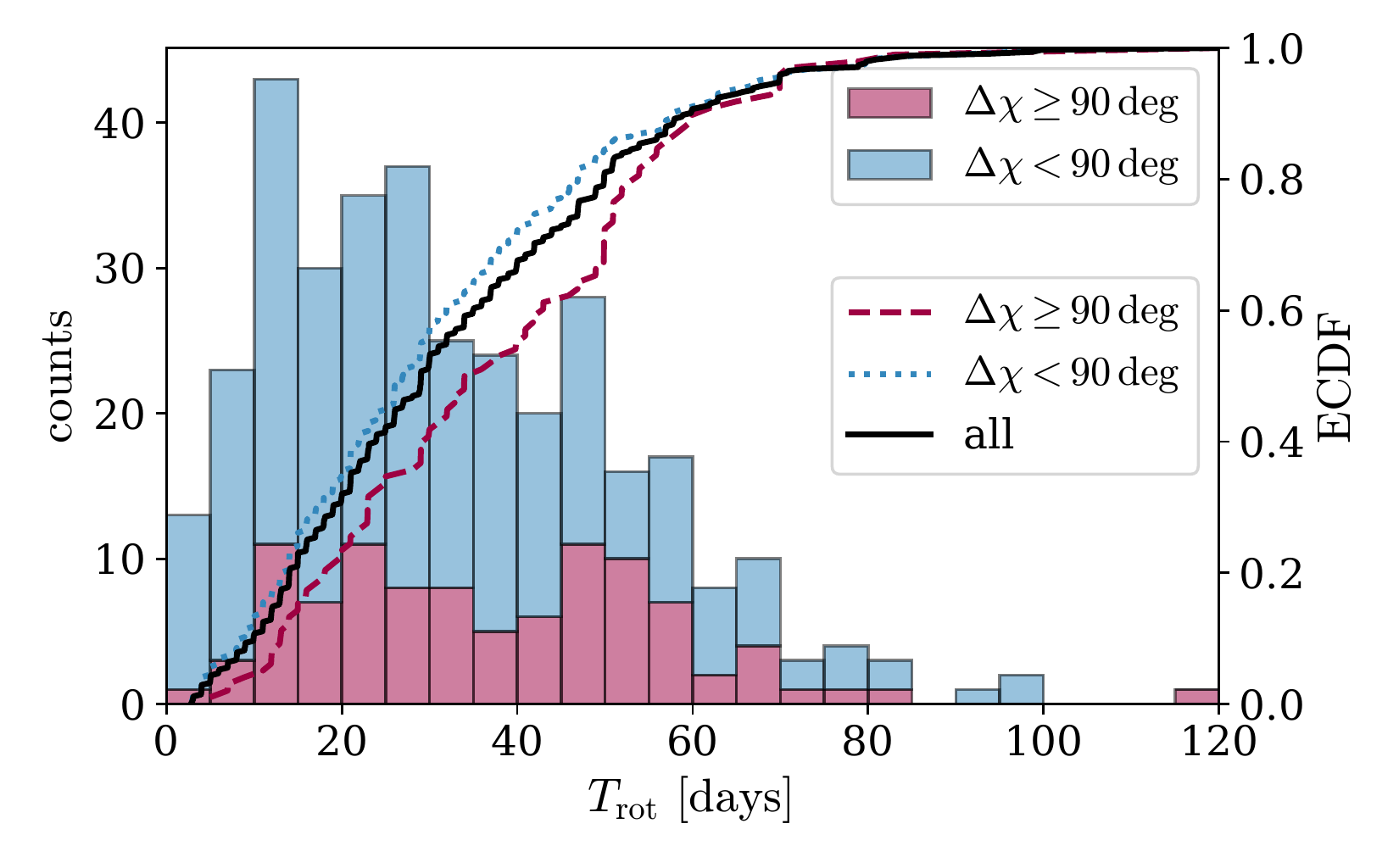}
  \caption{
    Distribution of rotation durations in the RoboPol main sample. The stacked 
    histograms are based on the large (red) and small (blue) rotations only, 
    the coloured ECDFs accordingly. The black ECDF is based on all rotations.}
  \label{fig:distdur}
\end{figure}

\begin{figure}
  \centering
  \includegraphics[width=\columnwidth]{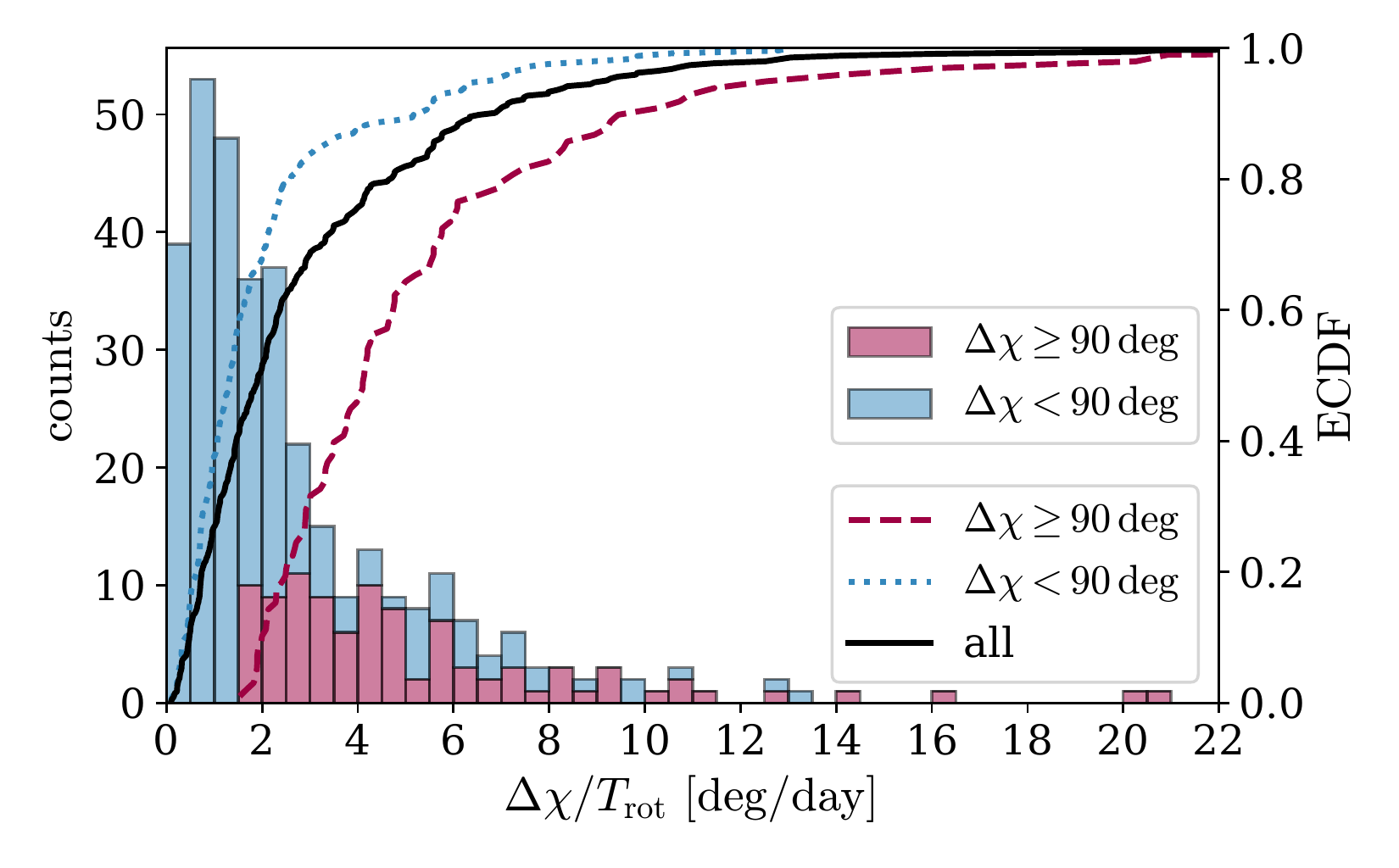}
  \caption{
    Distribution of rotation rates in the RoboPol main sample. The stacked 
    histograms are based on the large (red) and small (blue) rotations only, 
    the coloured ECDFs accordingly. The black ECDF is based on all rotations.}
  \label{fig:distrate}
\end{figure}

\begin{figure}
  \centering
  \includegraphics[width=\columnwidth]{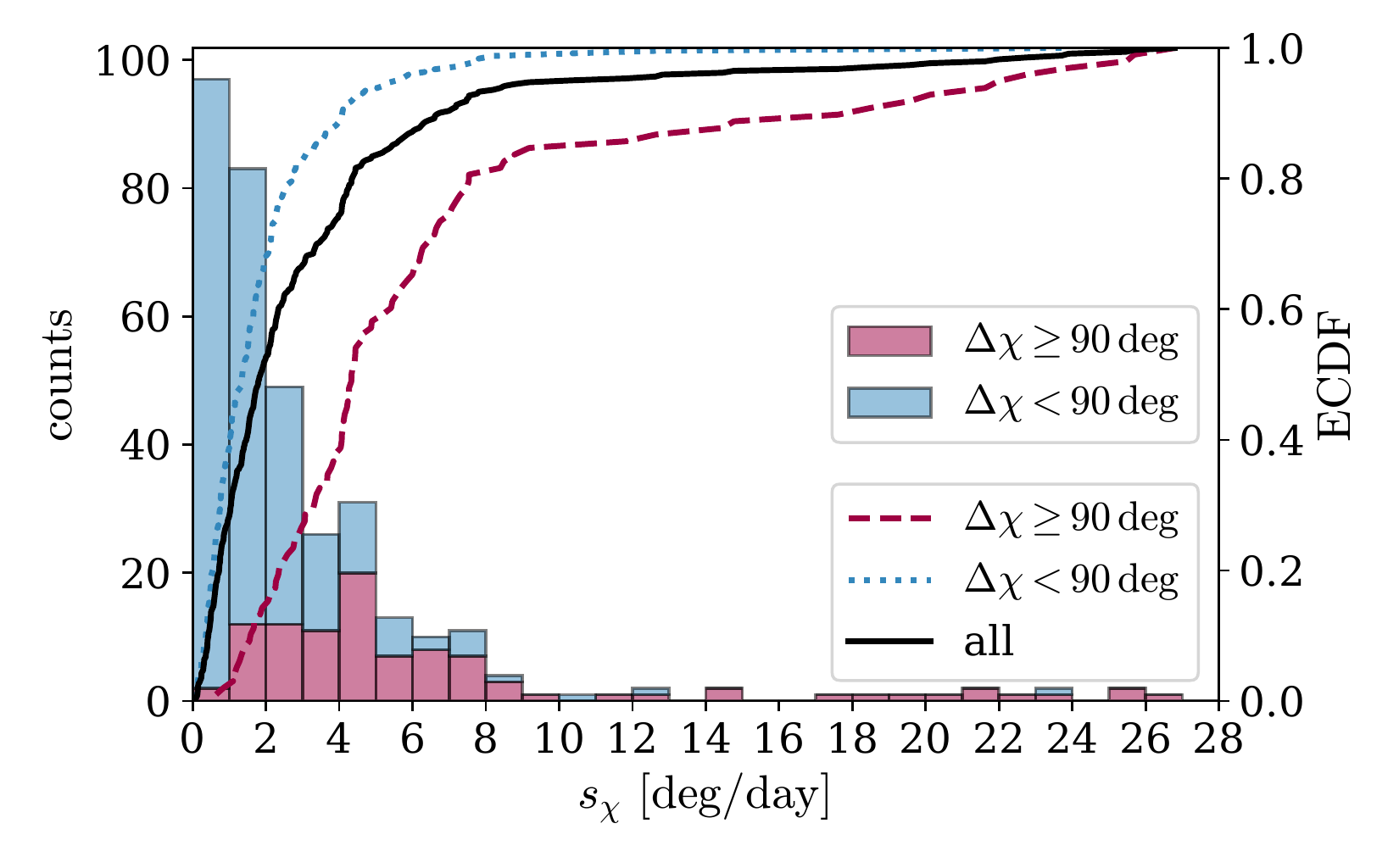}
  \caption{
    Distribution of the rotation variability in the RoboPol main sample. The 
    stacked histograms are based on the large (red) and small (blue) rotations 
    only, the coloured ECDFs accordingly. The black ECDF is based on all 
    rotations.}
  \label{fig:distvar}
\end{figure}

\Crefrange{fig:distampl}{fig:distvar} show the distributions of the four 
parameters characterizing the rotations that are found in the RoboPol main 
sample.
Minimum, maximum, mean, and median values of the distributions are given in 
\cref{tab:rotpar}.

The large rotations have on average slightly longer durations than the small 
rotations (\cref{fig:distdur}).
We use a two-sample Kolmogorov--Smirnov (KS) test to test whether the durations 
of the small rotations and the large rotations come from the same distribution.
This hypothesis cannot be rejected at the 1\%~significance level or 
lower (p-value: $0.01$), showing there is no significant difference between the
durations of small and of large rotations.
This implies that long-lasting rotations do not necessarily correlate with 
large rotation amplitudes.
Instead, it is mostly the rotation rate that should affect the amplitude.
Indeed, we find that the rotation rates of the population of large rotations 
are on average higher (\cref{fig:distrate}) and the KS~test indicates a 
significant difference between the two distributions (p-value:
$\sim 10^{-26}$).
Having the rotation amplitudes depend more on the rotation rate than on the 
rotation duration could be an artefact of the observations.
As the observing periods are limited, larger rotations have to be produced by
faster rotations.

$28\%$ of the measured amplitudes and durations are lower limits.
Therefore, the distributions shown in \cref{fig:distampl,fig:distdur} depend
on the sampling of the observations.
Likely, the intrinsic distributions are partially shifted to higher values.
The distribution of the average rotation rates are affected accordingly.
$28\%$ of the measurements may be over- or underestimated, if the rotation
rate during the rotations is variable.

Due to the $180\,\mathrm{deg}$~ambiguity, EVPA variability can only be measured 
correctly if the EVPA changes by less than $90\,\mathrm{deg}$ between two data
points.
Therefore, the median observation cadence of $6$~days corresponds to an upper 
limit for the detectable rotation rates of $\lesssim 15\,\mathrm{deg}/\mathrm{day}$.
The distribution of rotation rates is thus limited by the sampling cadence.
\Cref{fig:distrate} is cropped after the second fastest rotation with a rate
of $21\,\mathrm{deg}/\mathrm{day}$.
The fastest rotation with a rate of $52\,\mathrm{deg}/\mathrm{day}$ occurs in 
RBPL\,J2202+4216, showing a $259\,\mathrm{deg}$ rotation within five days, over 
five data points.
It is shown in Fig.~2 of \citet{2015MNRAS.453.1669B}\footnote{This rotation
is identically found with both methodologies used to identify rotations here 
and in \citet{2015MNRAS.453.1669B}.}.

\Cref{fig:distvar} shows the distribution of the EVPA variation estimator.
The two distributions based on the large and the short rotations are 
significantly different as indicated by a two-sample KS~test 
(p-value: $\sim 10^{-26}$).
The large rotations are on average more erratic than the small rotations.
The variation estimator is, per definition (cf.~Eq.~\ref{eq:varest}),
independent of the rotation rate, assuming a constant rate.
It is increased by erratic variability superimposed on a secular trend, 
non-linear trends, and measurement uncertainties, whereas the latter two are 
regarded as biases \citep{2016AA26A...590A..10K}.
Given the adaptive scheduling scheme of the RoboPol observations, large 
rotations are more densely sampled in time than small rotations, as stronger 
variability led to intensified monitoring.
Testing the two biases of the variation estimator for a dependence on the 
number of data points, we find that indeed both biases increase with larger 
numbers of data points.
The effect of this dependence is at least one order of magnitude lower than the 
difference between the two distributions shown in \cref{fig:distvar}.
Therefore, these biases cannot explain why large rotations are on average 
more erratic than small rotations.

In the following section we investigate potential relations between different 
rotation parameters taking the lower limit into consideration.

\subsubsection{Relations between the rotation parameters}
\label{sec:rotpardepobs}

\begin{figure*}
  \centering
  \includegraphics[width=\linewidth]{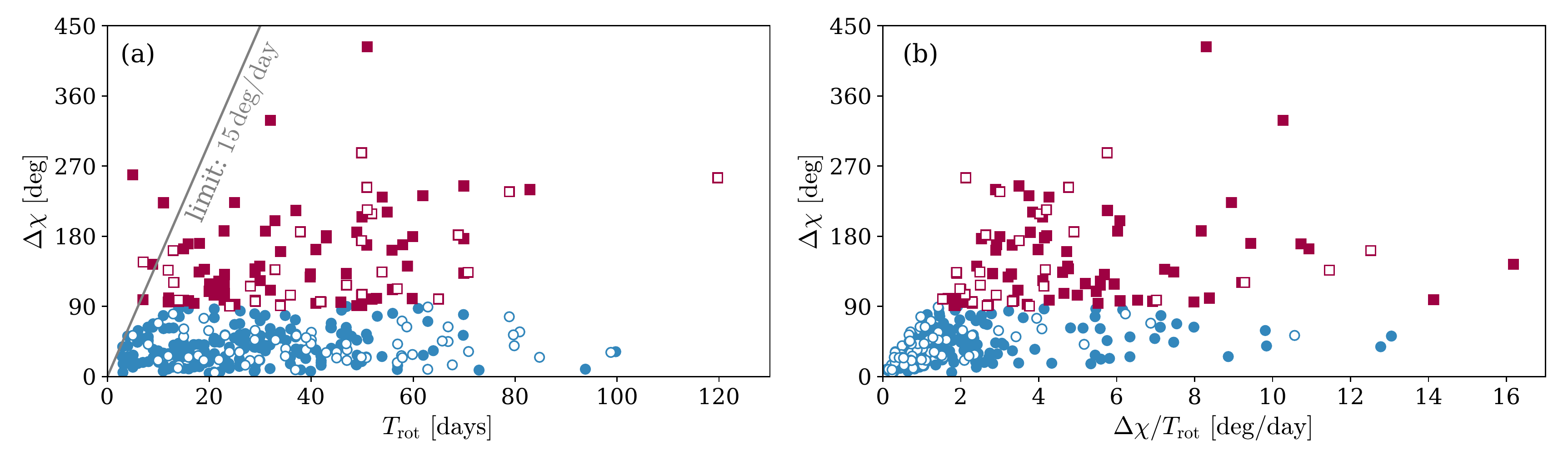}
  \caption{
    Dependence of the rotation amplitude on the rotation duration
    (\emph{panel~a}) and the rotation rate (\emph{panel~b}). Blue dots show 
    small ($\Delta\chi < 90\,\mathrm{deg}$), red squares large rotations 
    ($\Delta\chi \geq 90\,\mathrm{deg}$). Open symbols indicate lower limits
    for the amplitude and duration and uncertain rotation rates due to the 
    limited observing periods.
    The grey line marks the upper limit of measurable rotation rates, based on 
    the median time sampling.}
  \label{fig:amplvsdurandrate}
\end{figure*}

\Cref{fig:amplvsdurandrate} shows the rotation amplitude plotted against its 
duration and rate.
Small rotations are marked by blue dots, large rotations by red squares.
Open symbols mark the $28\%$ of the rotations that were limited by the 
observing periods.
These measurements are lower limits for the amplitudes and durations and 
uncertain for the rotation rate (cf.~\cref{sec:rotparbias}).

Rotations of any amplitude cover a large range of durations and rotation rates.
We use a generalized version of the Kendall rank correlation coefficient $\tau$
as described by \citet[][Chapter 10.4.1, Eq. 10.23]{FeigelsonBabu201208},
which accounts for censored data.
We treat the amplitudes and durations that are limited by the observing periods
as lower limits.
Rotation rates are treated as exact measurements as they can be either over- or
underestimated.
The correlation coefficients indicate that the rotation amplitude depends 
slightly more strongly on the rate ($\tau = 0.38$) than on the duration
($\tau = 0.32$).
When considering only the measurements and not the lower limits, the stronger
dependence of the rotation amplitude on the rotation rate ($\tau = 0.5$) rather
than on the duration ($\tau = 0.18$) becomes more evident.
These results are consistent with the discussion of the distributions in the 
previous section and indicate that the dependence of the amplitude on the rate
rather than the duration is an intrinsic effect and not a sampling artefact.

\begin{figure*}
  \centering
  \includegraphics[width=\linewidth]{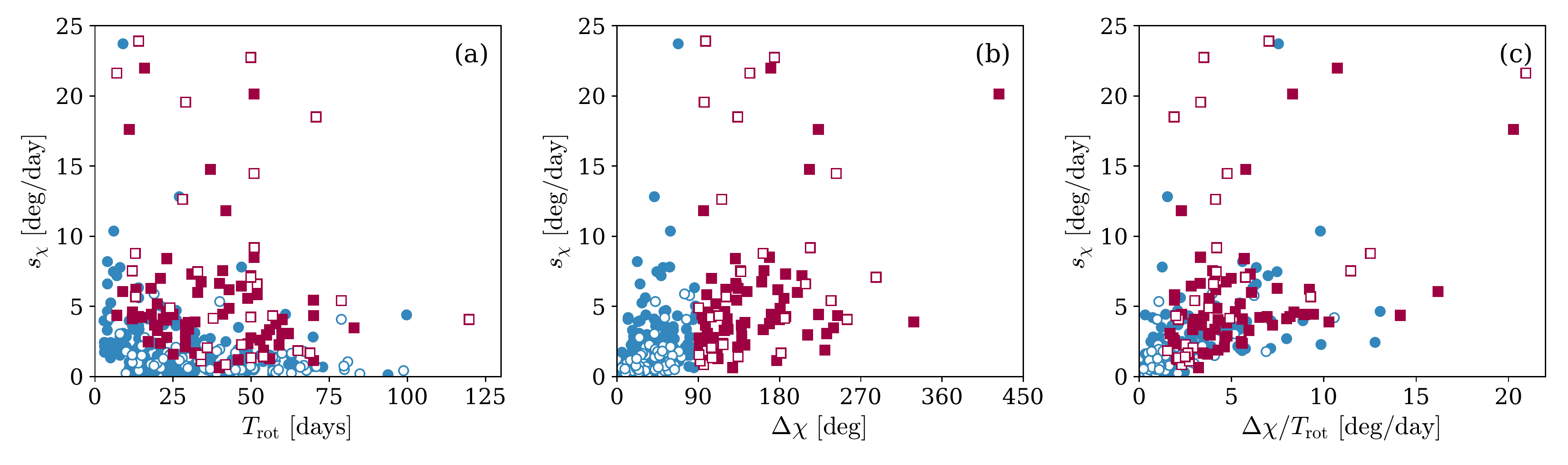}
  \caption{
    Dependence of the rotation variation estimator on the amplitude 
    (\emph{panel~a}), duration (\emph{panel~b}), and rate (\emph{panel~c}). 
    Symbols are as in \cref{fig:amplvsdurandrate}.}
  \label{fig:varvsrest}
\end{figure*}

\Cref{fig:varvsrest} shows the variation estimator plotted against the
rotation duration, amplitude, and rate.
We use the  generalized Kendall rank correlation coefficient $\tau$ to test
these pairs of parameters for correlations, considering the amplitudes and
durations limited by the observing periods as lower limits.
The correlation coefficient $\tau = -0.15$ does not indicate a strong relation 
between the variation estimator and the duration (panel~a).
On the other hand, $\tau=0.36$ for the amplitude and $\tau=0.56$ for the
rotation rate indicate a relation between these parameters and the variation 
estimator.
Thus, rotations with larger amplitudes are on average more variable, i.e., less
smooth.
Accordingly, higher rotation rates lead on average to stronger variability.

This behaviour would be expected from a stochastic process.
To produce larger rotations or higher rotation rates, stronger variability is 
needed in the EVPA, which increases the erratic behaviour of the rotations
\citep{2016AA26A...590A..10K}.
Consistently, strong erratic variability is less likely to produce a 
continuous rotation over a long period.
This behaviour is indicated in the negative sign of the weak correlation 
between variation estimator and duration (cf.~\cref{fig:varvsrest}, panel~a).
We test whether these rotations follow a stochastic process in 
\cref{sec:rotpardepobs}.

\subsection{Polarization fraction during rotation and non-rotation periods}
\label{sec:polfrac}

We test whether the polarization fraction behaves differently during rotations 
and non-rotation periods.
For each object we calculate the mean and standard deviation of the
polarization fraction during the rotation periods,
$\left< p \right>^\mathrm{rot}$ and $\sigma(p)^\mathrm{rot}$, based on all data
point that are part of a rotation period.
And we calculate these values, $\left< p \right>^\mathrm{non-rot}$ and
$\sigma(p)^\mathrm{non-rot}$, for the non-rotation periods, which are all data
points that are not identified as part of a rotation.
Then, we calculate the ratio of the mean polarization fraction during the
rotation and non-rotation periods, and the ratio of the standard deviation
accordingly.
We perform this analysis in two ways.
First, we consider \emph{all} rotations.
In the example of \cref{fig:rotationexample} all coloured data points (red and 
blue) contribute to $\left< p \right>^\mathrm{rot}$ and 
$\sigma(p)^\mathrm{rot}$ and black data points contribute to 
$\left< p \right>^\mathrm{non-rot}$ and $\sigma(p)^\mathrm{non-rot}$.
Second, we consider only \emph{large} rotations.
Here, only red data points contribute to $\left< p \right>^\mathrm{rot}$ and 
$\sigma(p)^\mathrm{rot}$, while black and blue data points contribute to
$\left< p \right>^\mathrm{non-rot}$ and $\sigma(p)^\mathrm{non-rot}$.

\begin{figure*}
  \centering
  \includegraphics[width=\linewidth]{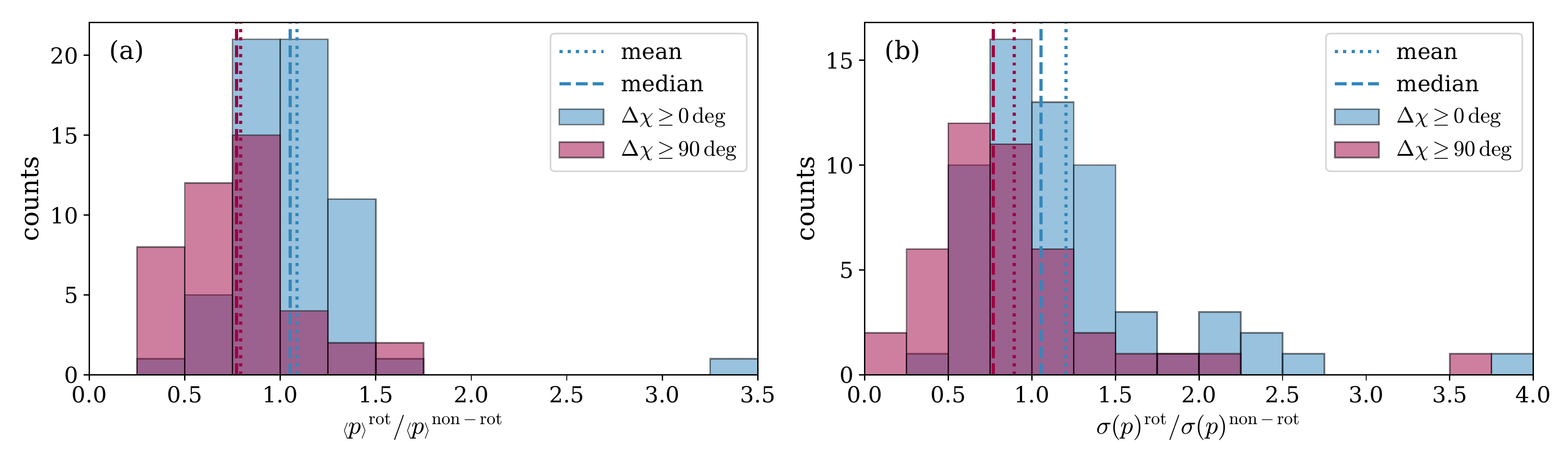}
  \caption{
    Distributions of the ratio between the polarization fraction mean during 
    rotations and during non-rotations (\emph{panel~a}) and the standard 
    deviation of the polarization fraction accordingly (\emph{panel~b}), based 
    on the RoboPol main sample. The blue histograms and  ECDFs consider all 
    rotations of arbitrary amplitudes, the red distributions accepted only 
    rotations of amplitudes $\Delta\chi \geq 90\,\mathrm{deg}$. The dotted and dashed 
    lines show means and medians of the corresponding distributions.}
  \label{fig:distpolratio}
\end{figure*}

The blue histograms and ECDFs in \cref{fig:distpolratio} show the distributions 
of the ratios of polarization means and ratios of polarization standard 
deviations, based on all identified rotations in the RoboPol main sample.
On average, the mean and standard deviation of the polarization fraction is
higher during the rotation periods than during the non-rotation periods.
The red distributions in \cref{fig:distpolratio} consider only large rotations 
as rotation periods.
In this case, the mean polarization fraction is on average lower during the
rotation periods than during the non-rotation periods and similarly the
standard deviation of the polarization fraction.
This comparison indicates that the large rotations exhibit a lower and less 
variable polarization fraction, whereas the small rotations mostly show a 
higher and more variable polarization fraction.
Similar results were reported in \cite{2016MNRAS.457.2252B}, where this 
analysis was based on maximum-likelihood estimates of the mean and the 
modulation index of the polarization fraction during rotations and 
non-rotations.


\section{Random walk simulations}
\label{sec:rwsim}

In \cref{sec:rotpardepobs} we argued that the observed relation between the 
rotation variability and the amplitude as well as the relation between the 
variability and the duration can be explained by a stochastic process.
In the following we test whether these data are consistent with a basic random 
walk process that emulates a turbulent flow in a turbulent magnetic field 
structure.

Turbulence may arise from current-driven instabilities 
\citep[e.g.][]{2012MNRAS.427.2480N} and transverse velocity gradients in the 
flow \citep[e.g.][]{2004ApJ...605..656V}.
\citet{2016AA26A...590A..10K} describe three basic random walk processes, using
cells that randomly change their properties to model the turbulence.
They show that the three processes cannot be distinguished from one another.
The only difference is the requirement of slightly different values for the
main model parameters to reproduce the same data.
Therefore, we limit our simulations and analysis to the ``simple $Q$, $U$
random walk process'', which we briefly outline in the following.

We create $N_\mathrm{cell}$ cells.
Each cell is assumed to contain a randomly oriented, uniform magnetic field.
Thus, the synchrotron emission from each cell is maximally polarized, 
$p_\mathrm{max} \approx 0.72$, assuming an electron energy spectrum with index
$2.5$ \citep[p. 217]{Longair201103}, and has a randomly oriented EVPA.
Additionally, we assume that each cell emits an equal amount of radiation.
Following \citet{2016AA26A...590A..10K}, we directly model the Stokes 
parameters $Q_i$ and $U_i$ for each cell $i$ through drawing two random numbers
$\hat{Q}_i \sim \mathcal{N}(0, 1)$, $\hat{U}_i \sim \mathcal{N}(0, 1)$ from 
a normal distribution.
The relative Stokes parameters are calculated from these random numbers as
\begin{align}
  q_i = \frac{Q_i}{I} \text{\ \ with\ \ }
  Q_i &= \frac{\hat{Q}_i}{\sqrt{\hat{Q}_i^2 + \hat{U}_i^2}} \cdot I_i \cdot m_{l, \mathrm{max}},
  \label{eq:normQ}  \\
  u_i = \frac{U_i}{I} \text{\ \ with\ \ }
  U_i &= \frac{\hat{U}_i}{\sqrt{\hat{Q}_i^2 + \hat{U}_i^2}} \cdot I_i \cdot m_{l, \mathrm{max}},
  \label{eq:normU}
\end{align}
which ensures the maximum polarization and equal intensity 
$I_i = \frac{I}{N_\mathrm{cells}}$ for all cells, where 
$I = \sum_{i=1}^{N_\mathrm{cells}} I_i$ is the total intensity.
As EVPA and polarization fraction are independent of the total intensity, it is 
set to $I=1$.
The polarization fraction $p$ and the EVPA $\chi$ are calculated from the 
summed Stokes parameters $q = \sum_{i=1}^{N_\mathrm{cells}} q_i$ and $u$ 
accordingly through
\begin{align}
  p &= \sqrt{q^2 + u^2}, \\
  \chi &= \frac{1}{2} \arctan \frac{u}{q},
\end{align}
using the \verb|atan2| definition of the arctangent that returns the correct 
quadrant of the angle in the range $(-180\,\mathrm{deg}, 180\,\mathrm{deg}]$.
At each time step several cells are randomly selected and their properties are 
changed randomly according to the previous formalism.
The cell variation rate $n_\mathrm{var}$, sets the number of cells that
change each unit time step.
We point out that we study and model the variability in the observer's frame.
The simulations are implicitly assumed to be affected by the same Doppler 
factor and redshift distribution as the data.
Modelled cell variation rates are not source intrinsic rates.

As discussed in \cref{sec:rotparbias} the parameters used to characterize 
rotations are affected by time sampling and measurement uncertainties.
For a reliable comparison of data and simulations the simulations need to be 
affected by the same biases.
Therefore, we simulate time steps and observational noise similar to the
observed data.
Random simulation time steps, $\Delta t_i$, are drawn from the ECDF of the
observation time steps (\cref{fig:distmodelpar}, panel~a) and simulation
durations, $T_\mathrm{sim}$, are randomly drawn from the ECDF of observing
periods (\cref{fig:distmodelpar}, panel~c).
We simulate observational errors through randomly drawing uncertainties 
$\sigma_{q,t}$ from the ECDF of the combined, observed uncertainties of $q$ and 
$u$ (\cref{fig:distmodelpar}, panel~c) for each simulation time $t$.
In the simulations we set $\sigma_{q,t} = \sigma_{u,t}$.
The random errors $q_{\mathrm{err},t}$, $u_{\mathrm{err},t}$ are drawn from a 
normal distribution $\mathcal{N}(0, \sigma_{u,t})$ and added to the summed
$q$ and $u$ for every time step before calculating the polarization fraction
and EVPA.

Before testing this model against the RoboPol main sample data in 
\cref{sec:simvsdata}, we discuss general dependencies on the model parameters
$N_\mathrm{cells}$ and $n_\mathrm{var}$.
Figure~5 in \citet{2016AA26A...590A..10K} depicts the relation between the
model parameters and the mean and standard deviation of the polarization 
fraction.
We limit the tested parameter space according to the polarization fraction 
observed in the RoboPol main sample.
The observed mean polarization fraction ranges from $1\%$ to $24\%$, the 
standard deviation from $0.6\%$ to $12\%$.
With 20~cells a mean polarization fraction of $\sim 15\%$ is expected,
though higher values may occur \citep[][Fig. 5]{2016AA26A...590A..10K}.
The lowest mean and standard deviation observed in the polarization fraction
may be produced by up to 1000~cells.
We test the following parameter space:

\begin{description}
  \item[Numbers of cells:] $N_\mathrm{cells} \in [20, 40,\dots, 1000]$,
  \item[Cell variation rates:] $n_\mathrm{var} \in [2, 4, \dots, 100]$~cells
       per day.
\end{description}

\subsection{Expected number of rotations}
\label{sec:exprotnum}

\begin{figure*}
  \centering
  \includegraphics[width=\linewidth]{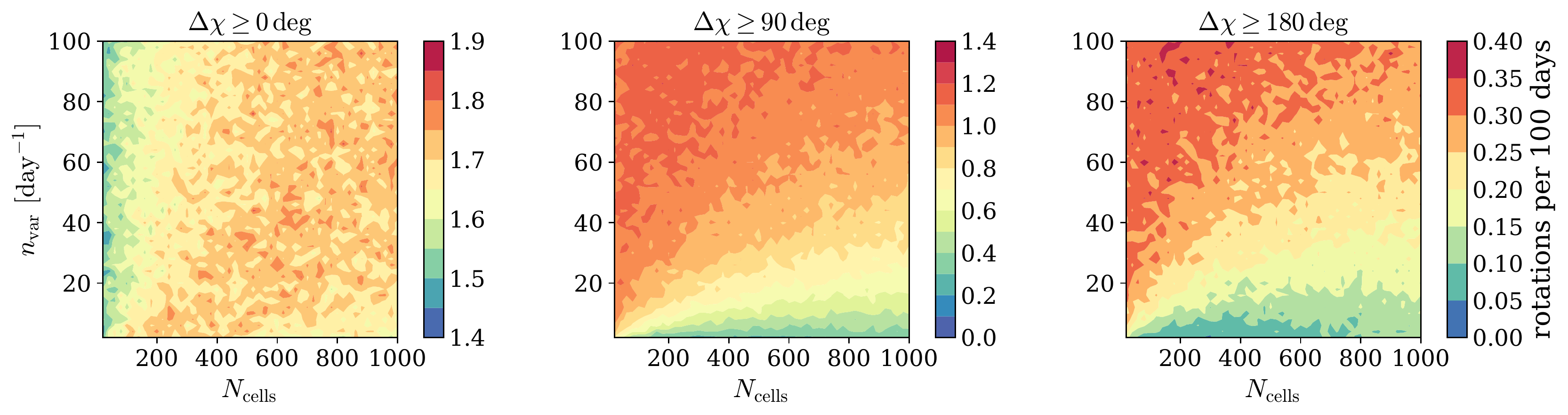}
  \caption{
    Expected number of rotations per 100~days with arbitrary amplitudes 
    (\emph{left~panel}), amplitudes larger than $90\,\mathrm{deg}$ (\emph{mid~panel}), and 
    amplitudes larger than $180\,\mathrm{deg}$ (\emph{right~panel}).}
  \label{fig:simrot}
\end{figure*}

For each parameter combination we run 500~simulations\footnote{With 50~test 
values for the number of cells and 50~values for the cell variation rate, we 
test 2500~points in the parameter space. Due to computer time restrictions we 
limit the number of simulations at each point to 500, which gives a total of 
1\,250\,000~simulations.} with total times $T_\mathrm{sim}$ randomly drawn 
from the ECDF of observing periods (\cref{fig:distmodelpar}, panel~b).
We count the identified rotations over the summed total time of each set
of 500~simulations.
\Cref{fig:simrot} shows the number of rotations per $100$~days expected from 
this random walk process depending on the model parameters for different 
minimum rotation amplitudes.

Considering rotations with arbitrary amplitudes, the left panel of 
\cref{fig:simrot} suggests that there is a ratio of number of cells and cell 
variation rates that increases the frequency of rotations.
Too many cells or too small variation rates decrease the frequency of rotations
as it becomes less probable to find significant variability in the EVPA.
On the other hand, very few cells or high variation rates decrease the frequency 
of rotations as the EVPA becomes more erratic.
The frequency of 1.57~rotations per 100~days observed in the RoboPol sample 
(\cref{tab:rotnum}) suggests that the sources should be located at small 
numbers of cells $N_\mathrm{cells} \lesssim 100$ in the model parameter space,
if they follow this random walk process.

Considering large rotations, the middle panel of \cref{fig:simrot} shows that 
lower variation rates and larger numbers of cells decrease the frequency of 
large rotations.
Both effects reduce the variability of the EVPA, which is needed to produce 
large rotations.
The frequency of 0.45~large rotations per 100~days in the RoboPol sample
(\cref{tab:rotnum}) indicates that these sources require 
$N_\mathrm{cells} \gtrsim 100$ and that approximately 2\% of the cells vary
per day.

The discrepancy between the location of the sources in the parameter space,
estimated on the basis of all rotations with arbitrary amplitudes or on the 
basis of large rotations, implies that not all sources can be located in the 
same region of the parameter space if they follow this model.

\subsection{Model parameter dependence}
\label{sec:parameterdependency}

We use the simulations described in \cref{sec:exprotnum} to estimate
expectation values for the rotation parameters -- amplitude, duration, rate,
and variability -- and investigate their dependence on the model parameters.

As for the frequency of rotations in \cref{fig:simrot}, we find similar 
dependencies between the rotation parameters and the model parameters.
Rotation amplitudes, rates, and variation estimators increase with fewer cells 
and higher cell variation rates, which increases the variability of the EVPA.
The longest rotation durations are obtained at a certain ratio between number 
of cells and cell variation rate.
Too low or too high variation rates decrease on average the rotation duration.
These dependencies show that the rotation parameters and particularly the 
combination of the rotation parameters allow sources to be located in the model 
parameter space.

In the tested parameter space we identify rotations with amplitudes up to 
$730\,\mathrm{deg}$, durations up to $194\,\mathrm{days}$, rates up to
$99\,\mathrm{deg}/\mathrm{day}$, and variation estimators up to $61\,\mathrm{deg}/\mathrm{day}$.
These values are not general results of the random walk process, but depend on 
the time sampling, observing periods, and EVPA uncertainties of the RoboPol
observations.
Nonetheless, they show that all rotation parameters observed in the RoboPol
main sample (\crefrange{fig:distampl}{fig:distvar}) can be produced by the 
random walk process within the tested model parameter space.

\subsection{Rotation parameter relations}
\label{sec:rotpardepsim}

In \cref{sec:rotpardepobs} we have discussed potential relations between the 
parameters that characterize the rotations in the RoboPol main sample.
We use the simulations described in \cref{sec:exprotnum} to test whether
these dependencies are consistent with the random walk process.
From the simulations run at a particular point in the model parameter space we
randomly select 343 rotations, the same number we have observed in the data.
We find that the random walk process can produce a set of rotations that 
occupy roughly the same rotation parameter space as in
\cref{fig:amplvsdurandrate,fig:varvsrest} and show similar relations between
the rotation parameters.
These simulations are not directly comparable to the data, as the RoboPol
sources may be located at different regions of the model parameter space.
Nevertheless, these results indicate that \emph{the relations between the
rotation parameters found in the data are consistent with a random walk 
process.}
We discuss a thorough test between data and simulations in 
\cref{sec:simvsdata}.

\subsection{Expected polarization fraction during rotations and non-rotations}
\label{sec:exppolfrac}

In \cref{fig:distpolratio} we have shown that the polarization fraction in
the RoboPol main sample is on average lower during large rotations than during
small and non-rotations.
Here we test whether in the random walk model the mean and the standard 
deviation of the polarization fraction change between periods that are 
perceived as rotations and non-rotations.

\begin{figure*}
  \centering
  \includegraphics[width=\linewidth]{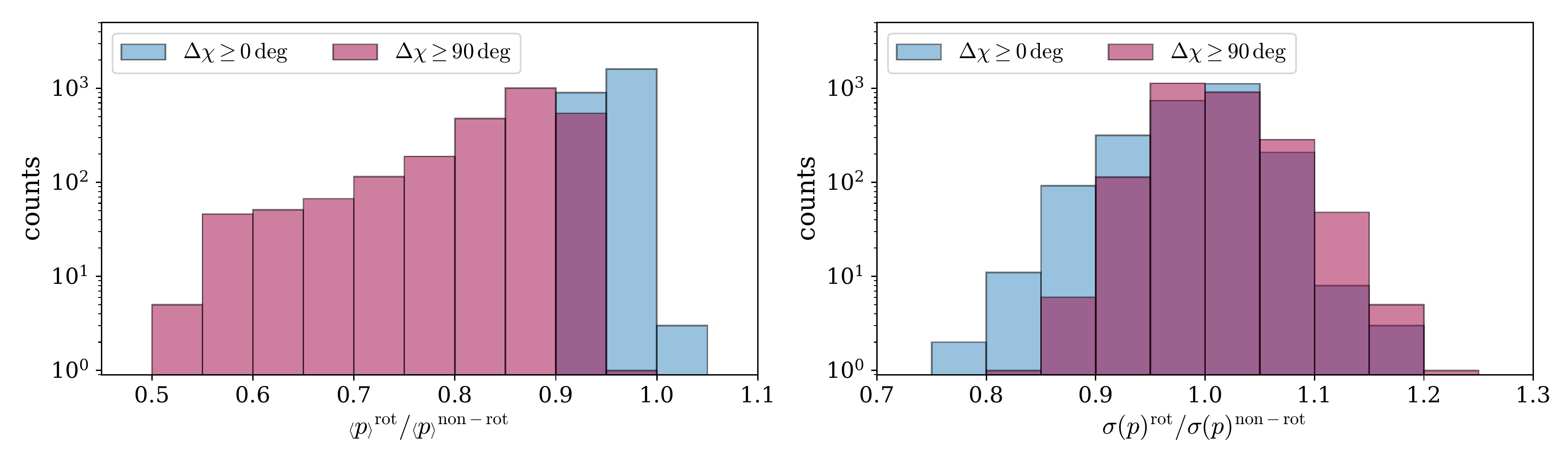}
  \caption{
    \emph{Left panel:} Distribution of the ratio of the mean polarization 
    fraction during rotations to non-rotation periods.
    \emph{Right panel:} Distribution of the ratio of the polarization fraction 
    standard deviation during rotations to non-rotation periods. The 
    distributions are based on all simulations covering the entire tested  
    parameter space with $T_\mathrm{sim} = 500\,000\,\mathrm{days}$.}
  \label{fig:simdistpolfull}
\end{figure*}

The data-based results are affected by sparse time sampling.
Rotation and non-rotation periods may consist of only a few data points, 
yielding unreliable estimates of the mean and standard deviation and the ratios
thereof.
In contrast, here we use reliable estimates by simulating long time series
with $T_\mathrm{sim} = 500\,000\,\mathrm{days}$.
For each parameter combination given in \cref{sec:rwsim} we run a long 
simulation with time steps randomly drawn from the distribution shown in 
panel~b of \cref{fig:distmodelpar}.
For each of the 2500~simulations we identify rotations and calculate the 
ratios of the mean polarization fraction (or standard deviation) during 
rotation and non-rotation periods, as described in \cref{sec:polfrac}.
The model results are shown in \cref{fig:simdistpolfull}.

The left panel shows the ratio of the mean polarization fraction during 
apparent rotations and non-rotation periods.
Considering only large rotations (red distribution), the mean polarization 
fraction is lower during apparent rotations than during non-rotations, ranging
down to half the mean polarization fraction of non-rotation periods.
Considering rotations with arbitrary amplitudes (blue distribution), the difference
in the mean polarization fraction is not as pronounced as for large rotations.
Still, the mean polarization fraction during apparent rotations is mostly below 
the mean polarization fraction of non-rotation periods.

An apparent EVPA rotation occurs when various cells vary their magnetic 
field orientation by chance, such that the net EVPA changes continuously in one 
direction.
In principle, a single, gradually changing cell can cause an apparent rotation,
when the polarization of all other cells roughly cancels out.
The polarization fraction is high when many cells accidentally align.
In this configuration, many cells are required to change accordingly to produce
an apparent rotations.
Thus, it is more likely that a few cells cause large rotations, while the 
other cells are unaligned and their polarization cancels out.
Consequentially, by picking out periods that show apparent large rotations we 
are more likely to select low polarized periods.
When selecting apparent rotations with arbitrary amplitudes, the difference in the 
mean polarization fraction between rotation and non-rotation periods becomes 
less pronounced for two reasons.
First, the fraction of data points selected as part of rotations with arbitrary
amplitudes is significantly higher than selecting only large rotations.
For example, $66\%$ of the RoboPol data points are part of rotations with 
arbitrary amplitudes, while only $22\%$ contribute to large rotations.
Therefore, the ratio of the mean polarization fraction is closer to one, when
considering all rotations.
Second, small rotations may occur with more cells aligned, i.e., higher 
polarization fraction, whereas it becomes less likely that many cells align 
subsequently to produce a large rotation.
Consequentially, small rotations are potentially more polarized than large 
rotations.

The right panel of \cref{fig:simdistpolfull} shows the ratio of the standard 
deviation of the polarization fraction during apparent rotations and 
non-rotation periods.
The standard deviation may be larger or smaller during apparent rotations 
than in non-rotation periods.
The distributions of the ratio peak around $1.0$ in both cases, considering
rotations with large amplitudes (red distribution) and arbitrary amplitudes (blue 
distribution).

By selecting periods of polarization data that apparently show a characteristic 
behaviour in the polarization angle, we select out periods that also differ in 
the polarization fraction compared to the rest of the data.
\emph{A difference in the polarization fraction between rotation and 
non-rotation periods does not necessarily indicate a different process during 
these periods, nor does it contradict a random walk process.}
In fact, it is consistent with a random walk process and is the result of a 
selection effect.
In particular, the observation that periods of large rotations in the RoboPol 
main sample are on average less polarized and less variable in the 
polarization fraction than during the non-rotation periods does not reject the 
random walk hypothesis.

The observed distributions in \cref{fig:distpolratio} and the simulation 
results in \cref{fig:simdistpolfull} are not directly comparable for two 
reasons.
First, we have simulated long time series with 
$T_\mathrm{sim} = 500\,000\,\mathrm{days}$ to avoid unreliable estimates of the 
mean and standard deviation of the polarization fraction.
Repeating the simulations with $T_\mathrm{sim}$ randomly drawn from panel~a of \cref{fig:distmodelpar}, the distributions shown in \cref{fig:simdistpolfull}
peak at value one, but extend to much higher values.
This shows that a higher mean polarization fraction during rotation periods is
possible, but is an artefact of a low number of data points.
Second, the distributions in \cref{fig:simdistpolfull} result from simulations 
at 2500~points in the model parameter space, equally distributed over the 
parameter space.
It is not known a~priori, where the RoboPol sources are located in the 
parameter space if they can be described by this random walk process.
A direct comparison of simulations and the RoboPol data is discussed in 
\cref{sec:simvsdata}.

\subsection{Mean polarization fraction and rotation rate}
\label{sec:polratecorrelation}

\citet{2016MNRAS.457.2252B} report a correlation between
$\left< p \right>^\mathrm{rot} / \left< p \right>^\mathrm{non-rot}$ and 
the logarithm of the source intrinsic rotation rate with a correlation 
coefficient $-0.66$ and slope $-0.19 \pm 0.07$, based on 16~sources for 
which Doppler factor and redshift were available.
We have shown in \cref{sec:exppolfrac} that large rotations in the random walk 
model are on average less polarized than non-rotation periods.
With a relation between the rotation rate and amplitude and a dependence 
between amplitude and the mean polarization fraction, the relation between 
$\left< p \right>^\mathrm{rot} / \left< p \right>^\mathrm{non-rot}$ and the 
rotation rate could be a result of a random walk process.

To test whether this relation is consistent with a random walk, we 
select one of the long simulations with $T_\mathrm{sim} = 500\,000\,\mathrm{days}$ 
described in \cref{sec:rwsim}, which covers one point in the model parameter 
space.
We randomly pick 16~rotations, estimate
$\left< p \right>^\mathrm{rot} / \left< p \right>^\mathrm{non-rot}$ and the 
rotation rate for each one, and run a linear regression over the mean 
polarization fraction ratio and the logarithm of the rotation rate.
Repeating this step for $20\,000$~iterations, we estimate the probability that
the linear regression yields a slope within the range $-0.19 \pm 0.07$ and a 
correlation coefficient $<-0.66$.
We repeat this analysis at various points in the model parameter space.
Depending on the position in the parameter space the probability of measuring
a similar slope as for the RoboPol data ranges between $14\%$ and $20\%$.
The probability of measuring a correlation coefficient more extreme than for 
the RoboPol data ranges from $1\%$ to $7\%$.

While the simulations are not directly comparable to the RoboPol data due to
the longer time-scale they cover and the unknown location of the RoboPol 
sources in the model parameter space, these results show that \emph{a relation 
between the mean polarization fraction ratio and the rotation rate does not 
reject the random walk hypothesis.}

\section{Testing simulations against data}
\label{sec:simvsdata}

In this section we discuss a method to constrain the model parameter space.
Then we test simulations against the data of the RoboPol main sample; 
focusing first on the characteristics of the EVPA rotations, second on the 
polarization fraction, and third on the observed number of rotations.

\subsection{Constraining the model parameter space}
\label{sec:minimization}

\begin{figure}
  \centering
  \includegraphics[width=\columnwidth]{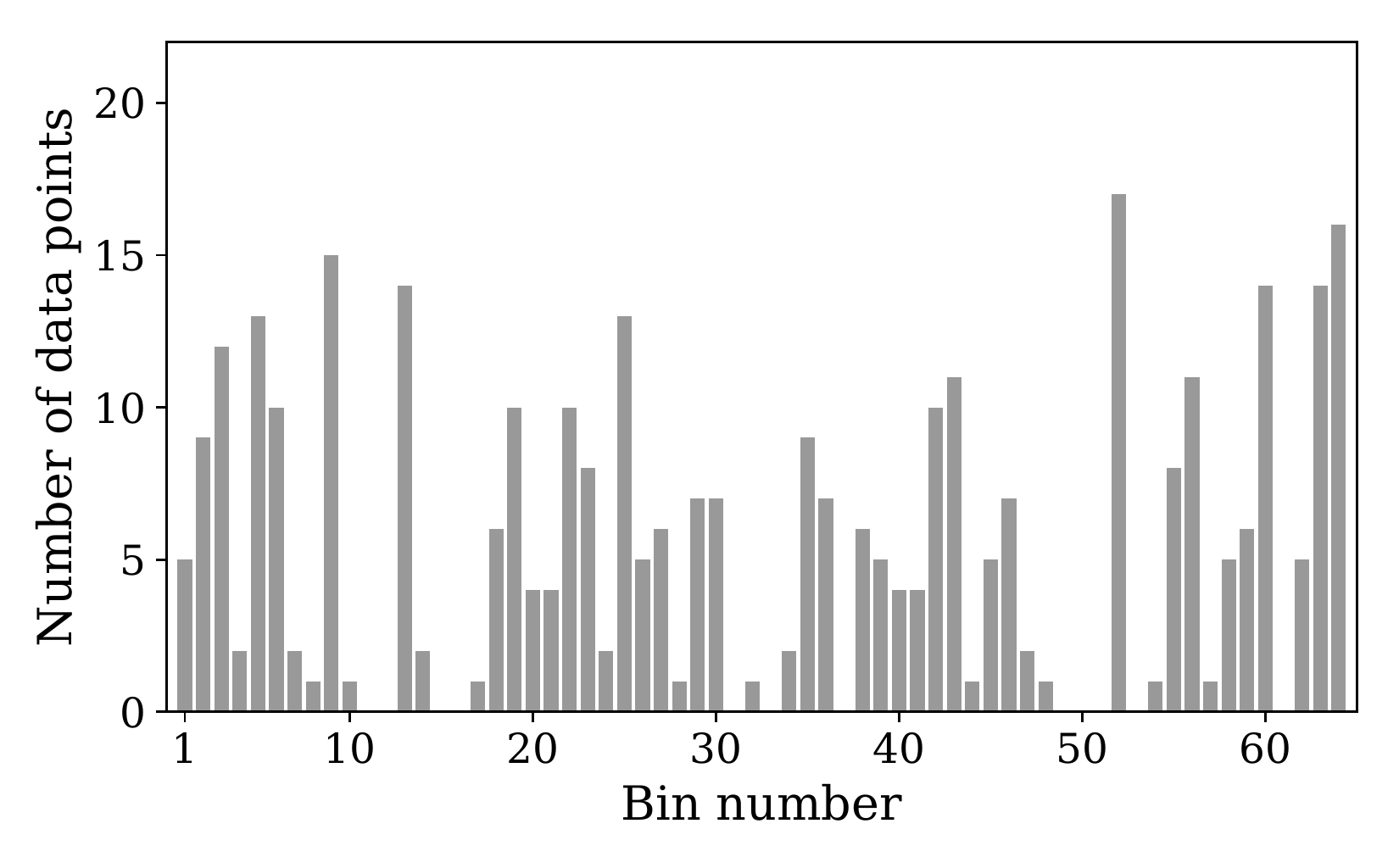}
  \caption{
    Number of data point contained in each bin of the rotation parameter space 
    amplitudened by amplitudes, durations, and variability.}
  \label{fig:binning}
\end{figure}

We characterize rotations by three independent parameters: amplitude, duration, and 
variation estimator.
We divide this three-dimensional parameter space by the following limits:
\begin{description}
  \item $\Delta\chi$: $0, 26, 49, 90, 423\,\mathrm{deg}$,
  \item $T_\mathrm{rot}$: $0, 16, 29, 47, 120\,\mathrm{days}$,
  \item $s_\chi$: $0, 0.8, 1.8, 4.1, 27\,\mathrm{deg}/\mathrm{day}$,
\end{description}
yielding four bins along each dimension and a total of 64 bins.
The bin ranges were chosen such that (i) most of the bins contain data points 
(52 out of 64) and (ii) the filled bins contain on average more than six data 
points.
The upper limits are based on the largest values observed in the RoboPol 
sample.
\cref{fig:binning} shows the number of data points for each bin, ranging 
between 1 and 17 for the filled bins.

In \cref{sec:parameterdependency} we have argued that the rotations resulting
from the random walk process depend on the number of cells $N_\mathrm{cells}$ 
and the cell variation rate $n_\mathrm{var}$.
These model parameters are a~priori unknown for all sources in the RoboPol 
main sample.
To compare our simulations against the observations, we need to limit the model
parameter space.
Therefore we assume that the number of cells and the cell variation rate of 
all sources follow predefined distributions, which are also a~priori unknown.
For the following explanation we assume that both parameters follow a 
log-normal distribution,
$N_\mathrm{cells} \sim \mathcal{LN}(M_c, S_c)$,
$n_\mathrm{var} \sim \mathcal{LN}(M_v, S_v)$,
with the corresponding distribution means $M_c$, $M_v$ and standard deviations 
$S_c$, $S_v$.
The log-normal distribution is a natural choice for both parameters as it gives
only positive values, but it is not physically justified.
We discuss alternative test-distributions later.
The distribution parameters $M_c$, $S_c$, $M_v$, $S_v$ are a~priori unknown.
We keep them as variables in the following minimization procedure:

\begin{enumerate}
\item Values for $M_c$, $S_c$, $M_v$, $S_v$ are randomly picked.

\item A random source sample is constructed picking 62~pairs of
model parameters, $(N_\mathrm{cells}, n_\mathrm{var})_i$, $i {=} 1, \dots, 62$,
from $\mathcal{LN}(M_c, S_c)$ and $\mathcal{LN}(M_v, S_v)$.

\item A sample of simulated EVPA curves is constructed. 
Here, we reproduce the same number of observing periods as in the RoboPol 
main sample (\cref{tab:obsperiods}), running four simulations for one 
object, three simulations for 58~objects, two simulations for two objects, and 
one simulation for the remaining object.
The total time $T_\mathrm{sim}$ of each simulation is randomly drawn from the 
ECDF of observing periods (\cref{fig:distmodelpar}, panel~a).

\item Rotations are identified in the simulation sample and the
corresponding rotation parameters are measured.
The rotations falling into each bin $N_{\mathrm{sim},i}$ are counted.

\item Steps~(ii) to~(iv) are repeated for $N_\mathrm{iter}$ iterations.
Each time the number of rotations per bin is stored.

\item For each bin $i$ we calculate the expected number of rotations
$\overline{N_{\mathrm{sim},i}}$ and the corresponding standard deviation 
$S_{\mathrm{sim},i}$ based on the $N_\mathrm{iter}$ iterations.
Then we calculate the statistics
\begin{align}
  \chi^2 (M_c, S_c, M_v, S_v) = 
  \sum\limits_{i=1}^{64} \frac{\left( \overline{N_{\mathrm{sim},i}} -
  N_{\mathrm{obs},i} \right)^2}{S_{\mathrm{sim},i}},
  \label{eq:chisq}
\end{align}
where $N_{\mathrm{obs},i}$ is the number of rotations of the RoboPol sample in 
each bin.
The sum is over all 64~bins.
\end{enumerate}

For step~(v) we choose $N_\mathrm{iter} = 500$.
The number of iterations is chosen relatively low to limit the calculation 
time, but large enough to get reliable estimates of 
$\overline{N_{\mathrm{sim},i}}$, $S_{\mathrm{sim},i}$, and 
$\chi^2 (M_c, S_c, M_v, S_v)$.

We run a minimization process over steps~(i) to~(vi) to find the optimal 
distribution parameters $M_c$, $S_c$, $M_v$, $S_v$ to produce a set of 
rotations comparable to the ones found in the RoboPol main sample.
For the minimization we use the \emph{Differential Evolution} algorithm 
by \citet{1997Storn} implemented in the \verb|scipy (v. 0.18.1)| python
package\footnote{\url{http://docs.scipy.org/doc/scipy/reference/generated/%
scipy.optimize.differential_evolution.html}}.

\subsubsection{Test distributions}
\label{sec:testdist}

As explained in the previous section, we assume that the number of cells and 
the cell variation rate follow a~priori unknown distributions.
We test various generic distribution types for both parameters:
(a) uniform distribution $\mathcal{U}(L,U)$ with lower limit $L$ and upper 
limit $U$, (b) log-normal distribution $\mathcal{LN}(M,S)$ with mean $M$ and 
standard deviation $S$, and (c) normal distribution 
$\mathcal{N}_\mathrm{trunc}(M,S)$ truncated such that all values are larger 
than zero, where $M$ and $S$ are the mean and standard deviation of the 
corresponding non-truncated normal distribution.
Values drawn for the number of cells are rounded up to the next integer.
For the cell variation rate we additionally test a truncated power-law 
distribution $\mathcal{PL}(\alpha,L,U)$ with index $\alpha$ and lower and upper 
limits $L$ and $U$.
A power-law distribution may be physically justified, assuming that the cell
variation rate corresponds to the flow speed and, thus, to the Lorentz factor,
which is found to follow a power-law distribution 
\citep{1997ApJ...476..572L}.
The cell variation rate would be additionally modulated by relativistic time 
effects and the source redshift.
Thus, modelling a direct correspondence between jet properties and the cell 
variation rate is beyond the scope of this simplistic modelling approach.
Therefore, we stick to these four generic distributions.

We test five combinations of distributions for the two model parameters as 
shown in \cref{tab:minimization}.
We run the minimization process described in \cref{sec:minimization} for each 
combination of distributions three times to check whether the process converges
at the same $\chi^2$ and the same best-fitting parameters.
We discuss the results of the minimization process in the following section.

\begin{table}
  \caption{
    $\chi^2$ estimates of the converged minimization process for five 
    combinations of test distributions.}
  \label{tab:minimization}
  \begin{tabular*}{\columnwidth}{@{\extracolsep{\fill} } r r r }
    \toprule
    \multicolumn{2}{c}{Test distribution}          & $\chi^2$  \\
    for $N_\mathrm{cells}$ &  for $n_\mathrm{var}$ &           \\
    \midrule
    uniform    & uniform    & $226 \pm 4$ \\
    normal     & normal     & $221 \pm 2$ \\
    log-normal & log-normal & $211 \pm 1$ \\
    log-normal & power-law  & $209 \pm 3$ \\
    normal     & power-law  & $205 \pm 1$ \\
    \bottomrule
  \end{tabular*}
\end{table}

An alternative approach to the minimization process described in 
\cref{sec:minimization} is to keep the two model parameters for each source as
variable in a Markov chain Monte Carlo procedure, minimizing the 
$\chi^2$-statistic of \cref{eq:chisq}.
This procedure does not require an assumption about the distributions of the 
number of cells and cell variation rates.
But with two parameters for 62~sources this procedure is a minimization over a
124~dimensional parameter space and computationally not feasible.

\subsubsection{Results}
\label{sec:minimizationresults}

\cref{tab:minimization} lists the minimized $\chi^2$, averaged over the three 
repetitions of the minimization process, and the corresponding standard 
deviation for the five different combinations of test distributions.
The minimized $\chi^2$ ranges between 205 and 226, showing that the 
best-fitting model does not strongly depend on the choice of the two model 
distributions.
The minimum $\chi^2$ is achieved with a truncated normal distribution 
$\mathcal{N}_\mathrm{trunc}(M_c,S_c)$ for the number of cells and a power-law 
distribution $\mathcal{PL}(\alpha_v,L_v,U_v)$ for the cell variation rate.
Therefore we limit all following discussion to this combination of test 
distributions.

\begin{figure}
  \centering
  \includegraphics[width=\columnwidth]{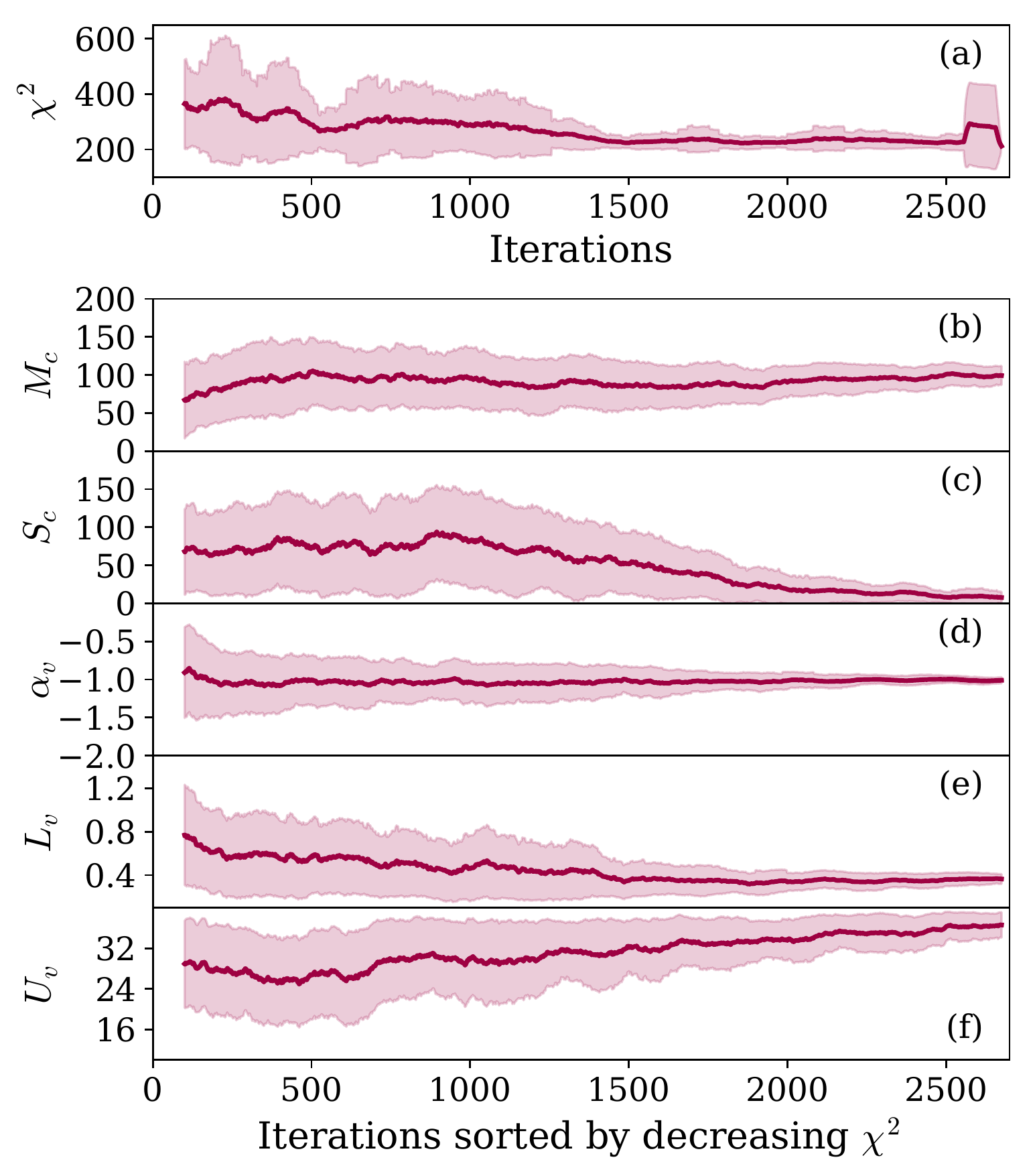}
  \caption{
    Progression of the $\chi^2$-minimization (panel~a). Each iteration refers 
    to one randomly picked set of parameters 
    $p = (M_c, S_c, \alpha_v, L_v, U_v)$. 
    These parameters are shown in panels~b--f. All parameter values are sorted 
    by decreasing $\chi^2$. In each panel, solid lines show the mean over the 
    previous 100~iterations, filled regions visualize the corresponding 
    standard deviation.}
  \label{fig:minimization}
\end{figure}

\Cref{fig:minimization} visualizes the progression of the differential 
evolution algorithm during one minimization process.
Every iteration represents one random pick of parameters\footnote{This set of 
parameters may or may not be accepted as potential solution by the algorithm.}
$p = (M_c, S_c, \alpha_v, L_v, U_v)$.
Panel~(a) shows the progression of $\chi^2 (p)$ over the iterations.
The solid line represents the mean over the previous 100~iterations and the 
filled area the corresponding standard deviation.
$\chi^2$ shows a decreasing trend and the spread is generally 
decreasing\footnote{Before accepting a solution the algorithm draws random 
picks throughout the parameter space to test, if it got stuck in a local 
minimum. These tests are indicated by the increased $\chi^2$ during the last 
$\sim200$~iterations.}, indicating that the minimization process is converging.

Panels~(b)--(f) show the progression of the five distribution parameters
$M_c$, $S_c$, $\alpha_v$, $L_v$, $U_v$ , sorted by decreasing $\chi^2$.
The solid line and the filled area illustrate the average and standard 
deviation over the previous 100~iterations.
All parameters are approaching a constant value and the spread starts to 
decrease at some point in the minimization progress, indicating good 
convergence.
The spread during the last iterations indicates how well the parameter is 
constrained.
The process shown in \cref{fig:minimization} converges at 
$M_c = 112$, $S_c = 2$, $\alpha_v = -1.0$, $L_v = 0.4$, $U_v = 38$ with 
$\chi^2 = 205$.
All repeated minimization processes converge at the same values within the
uncertainties indicated in \cref{fig:minimization}.

Under the assumption that all 62~sources of the RoboPol main sample follow the 
tested random walk process, the result of the minimization shows that all 
sources have roughly the same number of cells,
$N_\mathrm{cells} \sim 112 \pm 2$, and the cell variation rates range between
$\sim0.4$ and $38$~cells per day, whereas lower rates are more likely than 
higher rates.
%
With 63~degrees of freedom $\chi^2 = 204$ refers to a p-value of 
$\sim 10^{-16}$.
Based on this p-value the hypothesis that the observed rotations are produced 
by the random walk process with objects picked from the parameter space 
described by $(M_c, S_c, \alpha_v, M_v, S_v)$ is therefore rejected at high 
significance.
Since several bins contain only a few data points (cf.~\cref{fig:binning}),
this $\chi^2$-test is not a reliable hypothesis test.
Therefore, we use the $\chi^2$-minimization just to constrain the 
$(N_\mathrm{cells}, n_\mathrm{var})$ parameter space and run additional tests 
in the following section, based on the optimized parameter space found here.

\subsection{Simulated rotation samples}
\label{sec:simsamples}

With $(M_c, S_c, \alpha_v, L_v, U_v)$ fixed by the previous minimization 
process, we create random samples of rotations following steps~(ii) and~(iii) 
in \cref{sec:minimization}.
Then, the rotations are identified in the simulated data set and the 
distributions of the resulting rotation parameters are compared to the observed
ones (cf. \crefrange{fig:distampl}{fig:distvar}) with a two-sample 
Kolmogorov-Smirnov (KS) test.

\begin{figure*}
  \centering
  \includegraphics[width=\linewidth]{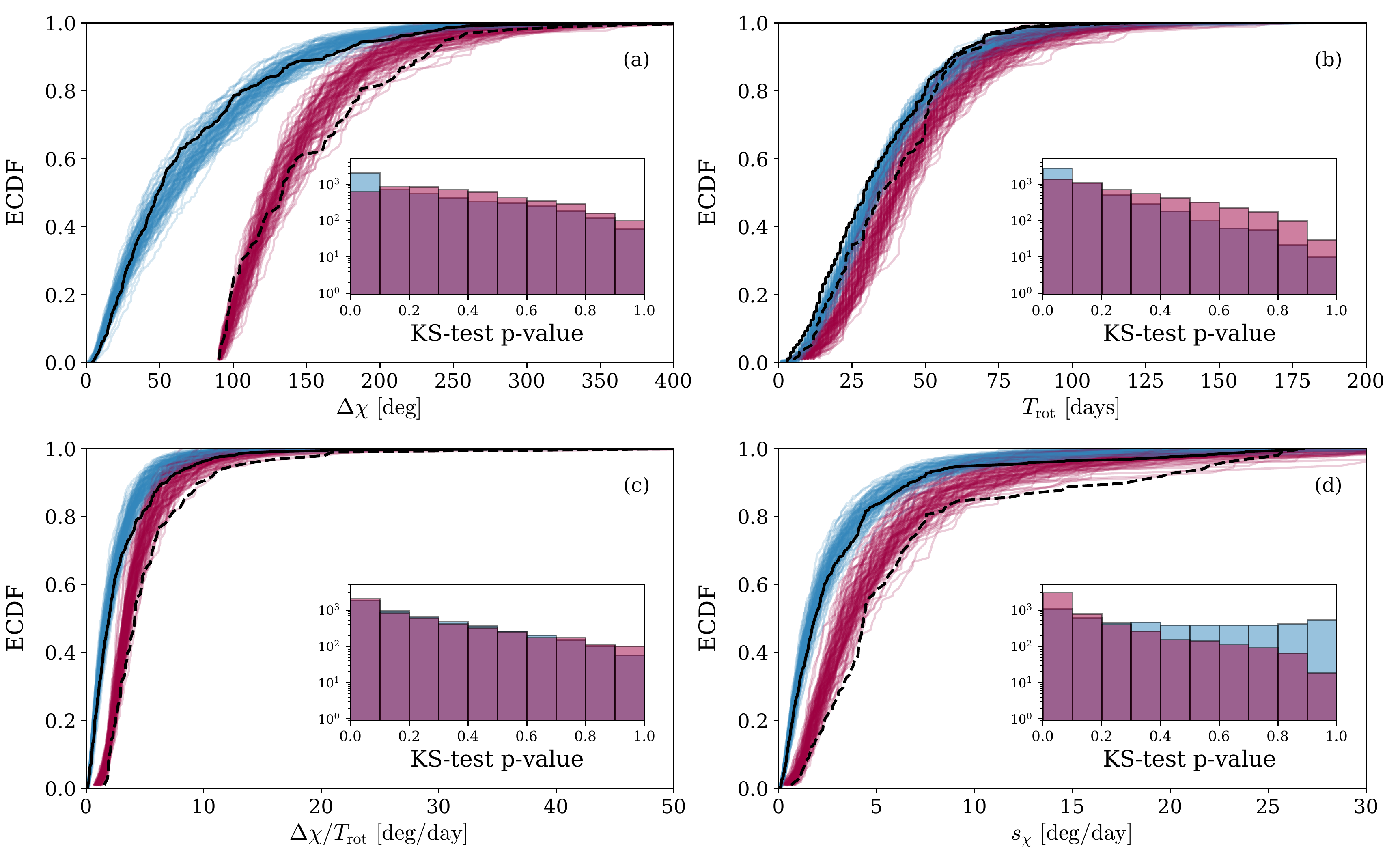}
  \caption{
    Cumulative distribution functions of rotation amplitudes (\emph{panel~a}), 
    durations (\emph{panel~b}), rates (\emph{panel~c}), and variability 
    (\emph{panel~d}). Coloured ECDFs are results of the random walk 
    simulations: blue ones are based on rotations with arbitrary amplitudes, 
    red  ones on rotations larger than $90\,\mathrm{deg}$. Each coloured ECDF 
    corresponds to one of 500~simulations. Solid and dotted, black lines are 
    the corresponding distribution functions based on the RoboPol main sample.
    Inset histograms show the distribution of p-values resulting from the 
    KS-tests comparing the blue simulation ECDFs with the observed, solid 
    black line ECDF.}
  \label{fig:simrotecdfs}
\end{figure*}

We run 5\,000~simulations, each producing a data set with the same number of
sources and periods per source as well as similar period durations, time
sampling, and uncertainties as the RoboPol data.
Each coloured line in \cref{fig:simrotecdfs} is based on one simulation, 
showing the ECDF of the rotation amplitudes (panel~a), durations (panel~b), 
rates (panel~c), and variation estimator (panel~d).
Red lines are based on large rotations, whereas blue lines include all 
rotations with arbitrary amplitudes.
The ECDFs are shown for 100~simulations.
Black lines (dashed and solid) show the corresponding observed distributions.

The inset histograms show the distributions of p-values resulting from the
two-sample KS-tests comparing each simulated distribution to the corresponding 
observed distribution of rotation parameters.
Blue histograms correspond to the distributions based on all rotations (blue
ECDFs and black solid line), red histograms to the distributions based only 
on the large rotations (red ECDFs and black dashed line).
The hypothesis that the compared distributions come from the same parent 
distribution, i.e., that the observed rotations originate from the tested
random walk model, cannot be rejected for any of the rotation parameters at
any reasonable significance level ($\alpha \leq 1\%$).
When we consider all three independent parameters characterizing the rotations
-- amplitude, duration, and variation estimator -- simultaneously, 
$74\%$ of the simulations are not rejected in any of the three KS-tests 
at $1\%$-significance level, based on all rotations.
Considering only large rotations, $77\%$ of the simulations are not 
rejected.

Whereas the KS-tests do not reject the hypothesis that the entire sample of 
rotations in the RoboPol main sample is produced by the tested random walk
process, a direct comparison of the observed and simulated ECDFs of the 
rotation parameters in \cref{fig:simrotecdfs} shows that the distribution of
parameters is not exactly reproduced.
Comparing first the distributions including all rotations (black solid and blue
lines), the distribution of rotation amplitudes (panel~a) is re-constructed 
well by the random walk process.
But simulated rotation durations (panel~b) are mostly larger and rotation rates 
(panel~c) accordingly slower than the observed ones.
$38\%$ of the simulated duration ECDFs lie entirely right of the 
observed ECDF, and $21\%$ of the rotation rate ECDFs lie entirely left
of the observed one\footnote{In this analysis we exclude the 20~smallest and 
20~larges values.}.
The ECDFs of the variation estimator (panel~d) indicate that the simulated
rotations are on average less variable than the observed ones.

A comparison between the simulated and observed ECDFs based on only the large 
rotations (black dashed and red lines) gives similar results and the observed 
and simulated ECDFs differ even more.
The reason is that the minimization process described in 
\cref{sec:minimization} that tries to find the optimal model parameter space
uses all rotations, which are dominated by small rotations (\cref{tab:rotnum}).
Therefore, the minimization process is more likely to match the ECDFs of all 
rotations than the large rotations.
This could indicate that the variability process responsible for the small 
rotations could be explained more easily with a random walk process, while the 
large rotations are less likely to follow a random walk process.
The amplitude cut between two different processes is not necessarily at a value
of 90~degrees, though, or clearly fixed at all.
Running this analysis only on a sub-sample of rotations -- small and large ones
individually -- is currently not possible, because the number of rotations is 
not large enough.
Therefore, we restricted this model test to trying to explain the entire sample
of rotations with the same process, without applying an arbitrary amplitude
cut.

\subsection{Polarization fraction of random walks}
\label{sec:simpolfrac}

With $(M_c, S_c, \alpha_v, L_v, U_v)$ fixed by the minimization process, we create 
samples of random walk simulations as in the previous section, but here we 
focus on the polarization fraction.
For each simulated object we estimate the mean and standard deviation of the 
polarization fraction during rotation periods, non-rotation periods and in 
general.

\begin{figure*}
  \centering
  \includegraphics[width=\linewidth]{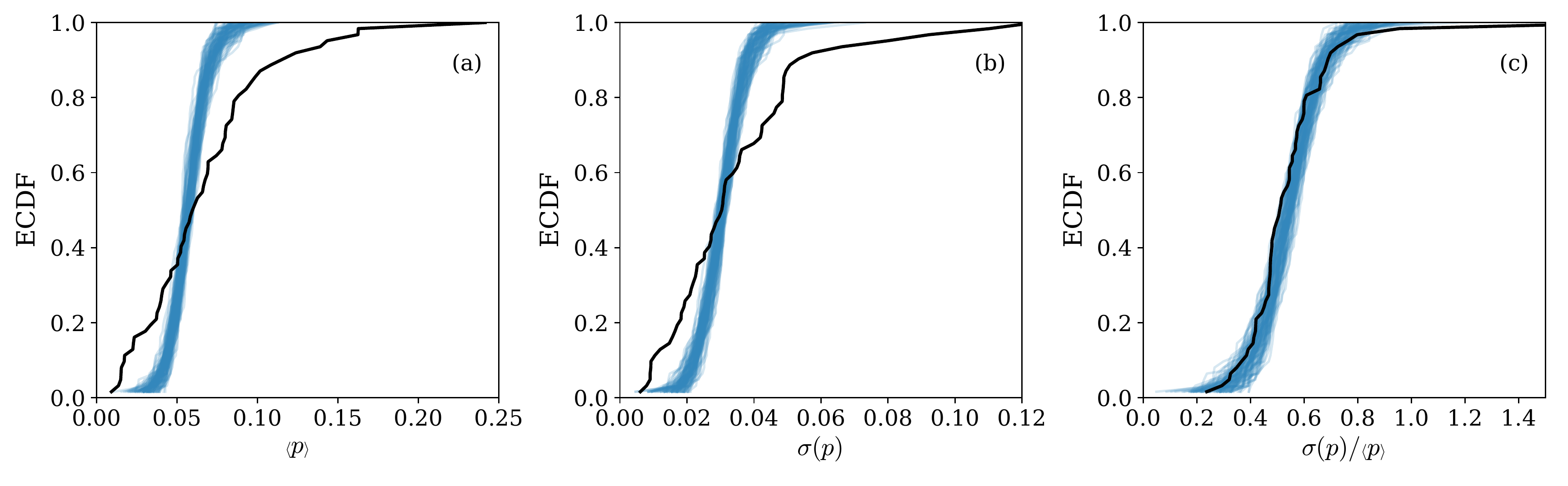}
  \caption{
    Cumulative distribution functions of the mean (\emph{panel~a}) and the 
    standard deviation (\emph{panel~b}) of the polarization fraction. Coloured 
    ECDFs are results of the random walk simulations. Black lines are the 
    corresponding distribution functions based on the RoboPol main sample.}
  \label{fig:simpolecdfs}
\end{figure*}

\Cref{fig:simpolecdfs} shows ECDFs of the mean (panel~a), the standard
deviation (panel~b), and the ratio of the two for 100~out of 5000~random 
samples (blue lines) in comparison to the distributions estimated from RoboPol 
main sample (black lines).
The simulated and observed distributions of the mean and standard deviation of
the polarization fraction differ significantly.
Two-sample KS-tests give p-values in the range of $10^{-5}$--$10^{-1}$ for the 
mean polarization fraction, rejecting $85\%$ of the simulations at 
$1\%$-significance level, and $10^{-4}$--$10^{-1}$ for the standard deviation 
of the polarization fraction with a rejection rate of $54\%$.
A direct comparison shows that the range of the simulated mean and standard 
deviation of the polarization fraction is much narrower than in the observed
data.
The origin of this narrow distribution is the small range of cell numbers that 
has been found by the minimization process in \cref{sec:minimization}, which 
the simulated sources are drawn from.
As shown in Fig.~5 of \citet{2016AA26A...590A..10K} the mean and standard 
deviation of the polarization fraction correspond to the number of cells 
in the random walk model and are thus limited by the allowed range of cell 
numbers.

Despite this result, the ratio of the standard deviation and the mean is 
comparable to the observations.
At $1\%$-significance level only $2\%$ of the simulations are rejected.
The reason is that this ratio typically is close to a value of $0.5$ in the 
random walk simulations as discussed in \citet{2016AA26A...590A..10K}, which is
what we also observe in the data.

\begin{figure*}
  \centering
  \includegraphics[width=\linewidth]{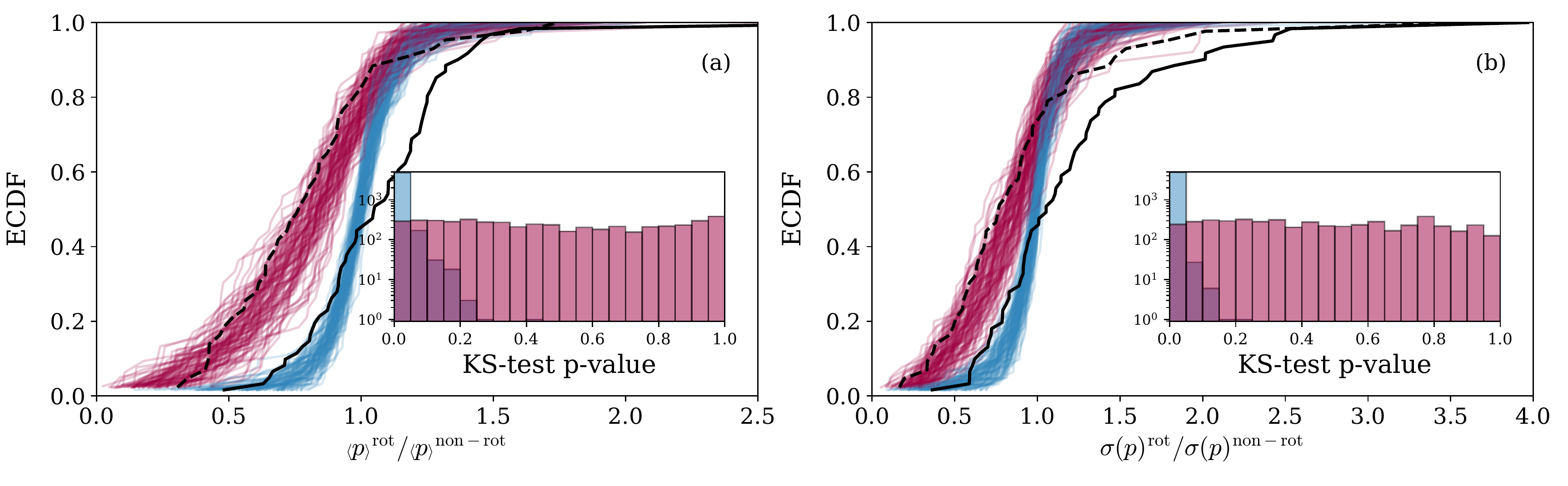}
  \caption{
    Cumulative distribution functions of the ratio of the mean polarization 
    fraction comparing the rotation periods to the non-rotation periods 
    (\emph{panel~a}) and the ratio of the the standard deviation 
    (\emph{panel~b}) accordingly. Coloured ECDFs are results of the random walk 
    simulations: blue ones are based on identified rotations with arbitrary 
    amplitudes, red ones on rotations larger than $90\,\mathrm{deg}$. Each coloured 
    ECDF corresponds to one of 500~simulations. Solid and dotted, black lines 
    are the corresponding distribution functions based on the RoboPol main 
    sample. Inset histograms show the distribution of p-values resulting from
    the KS-tests comparing the blue simulation ECDFs with the observed, solid 
    black line ECDF.}
  \label{fig:simpolratioecdfs}
\end{figure*}

The ratio between the mean polarization fraction during large rotations
and during non-rotation periods as well as the corresponding ratio of standard
deviations as observed in the data is reproduced by the simulations
(\cref{fig:simpolratioecdfs}, black dashed and red lines), showing that a
drop of the mean polarization fraction during large rotations is consistent
with a random walk process.
This is not the case when considering rotations with arbitrary amplitudes 
(black solid and blue lines), where $95\%$ of the simulations are 
rejected at $1\%$-significance level based on two-sample KS-tests.
The simulated distributions are much narrower than the observed ones, owing to 
the narrow distributions in the mean and standard deviation of the polarization 
fraction (\cref{fig:simpolecdfs}).

\subsection{Likelihood of observing no rotations}
\label{sec:simnorot}

\begin{figure*}
  \centering
  \includegraphics[width=\linewidth]{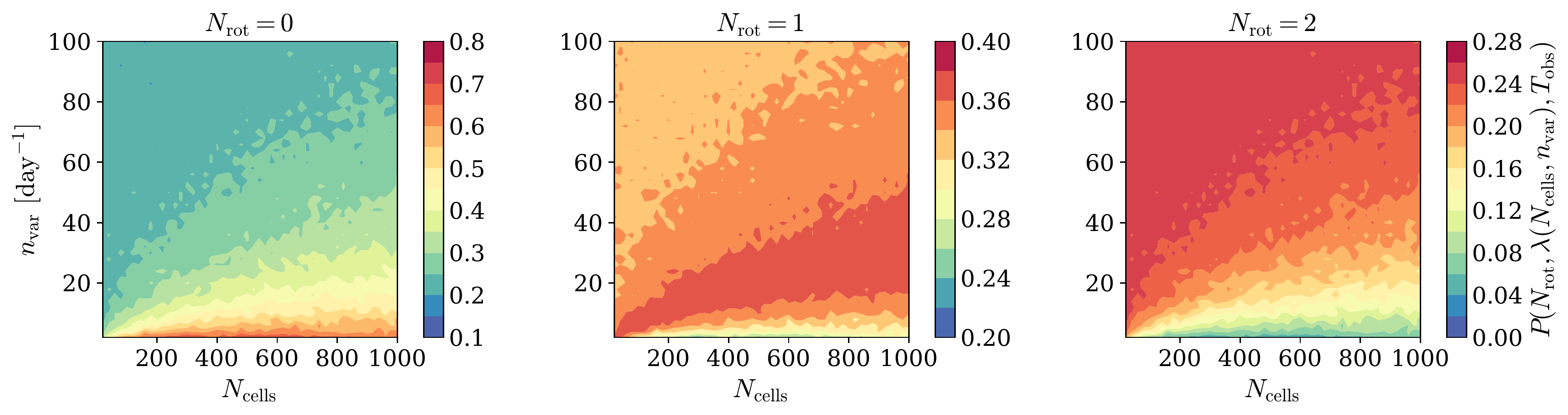}
  \caption{
    Probability of observing zero (\emph{left panel}), one (\emph{mid panel}), 
    or two (\emph{right panel}) rotations within the median  period 
    of the RoboPol sample, $T_\mathrm{obs} = 135$~days, from the tested random 
    walk process.}
  \label{fig:simrotprob}
\end{figure*}

\citet{2016MNRAS.462.1775B} argue that there are probably two 
classes of blazars in the RoboPol main sample: those which exhibit large 
($\Delta\chi \geq 90\,\mathrm{deg}$) EVPA rotations ($\sim 25\%$ of the sources) 
with an average frequency of one per $230$~days, and those which do not or 
rarely show rotations with a estimated frequency of less than one per
$2900$~days.
We note that in this study, as we use a less conservative method to identify
rotations, 43 out of 62 sources ($69\%$) show at least one large 
rotation.
We estimate the probability of observing $N_\mathrm{rot}$ rotations 
within time $T$ under the hypothesis that the polarization variability is 
produced by the tested random walk process, using the Poisson distribution
\begin{align}
  P(N_\mathrm{rot}, \lambda, T) 
  = \frac{(\lambda T)^{N_\mathrm{rot}}}{N_\mathrm{rot}!} 
  \mathrm{e}^{-\lambda T},
\end{align}
where $\lambda$ is the average frequency of rotations.
We use the frequency of rotations with amplitudes $\Delta\chi \geq 90\,\mathrm{deg}$
as shown in the mid panel of \cref{fig:simrot} as estimates of $\lambda$, 
depending on the model parameters $N_\mathrm{cells}$ and $n_\mathrm{var}$.
\Cref{fig:simrotprob} shows the probability of observing zero, one, or two 
rotations in the median RoboPol  period, $T_\mathrm{obs} = 135$~days, 
over a range of model parameters.
Observing one large rotation is generally more likely than observing no large 
rotation throughout the tested parameter space.
In the optimal parameter space the probability of observing no large rotation 
within a period of $T_\mathrm{obs} = 135$~days ranges between $23\%$ and 
$60\%$, consistent with the $33\%$ of RoboPol sources not showing large
rotations.
We note, though, that we do not account for multiple observing seasons in this
comparison.
This is considered in the following test.

\begin{figure}
  \centering
  \includegraphics[width=\columnwidth]{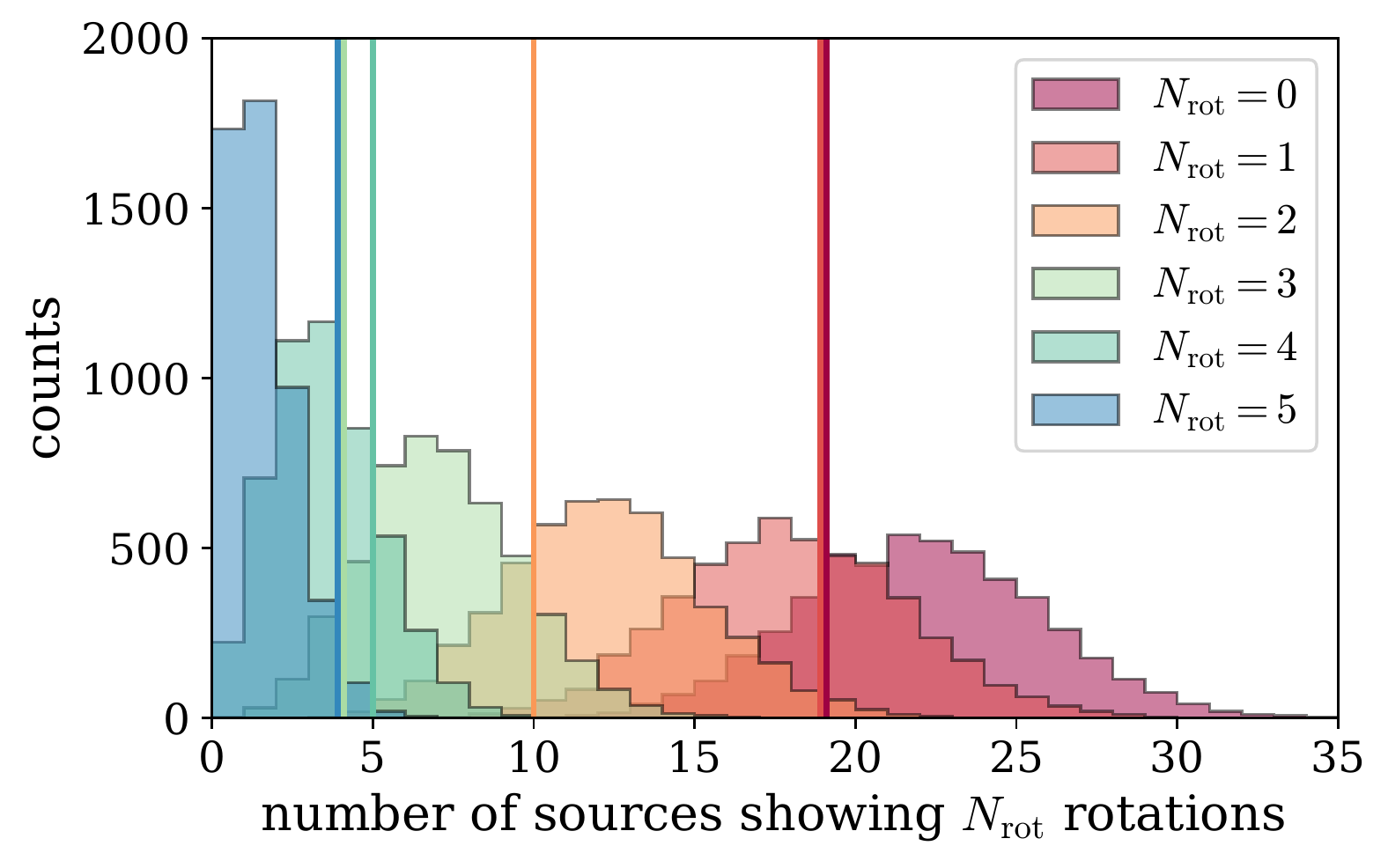}
  \caption{
    Distribution of the number of sources showing $N_\mathrm{rot}$ large 
    rotations based on 5000~simulation samples. The number of sources showing 
    $N_\mathrm{rot}$ large rotations in the RoboPol sample are marked by 
    vertical lines in corresponding colours.}
  \label{fig:simrotpersource}
\end{figure}

As in the previous sections, we construct simulated source samples.
Here, we count the number of sources showing $N_\mathrm{rot} = 0, 1, 2, \cdots$ 
rotations with large amplitudes, $\Delta\chi \geq 90\,\mathrm{deg}$.
Based on 5000~simulations, \cref{fig:simrotpersource} shows distributions of
the number of sources with $N_\mathrm{rot}$ rotations.
The number of sources showing $N_\mathrm{rot}$ in the RoboPol sample are marked
by vertical lines and are consistent with the simulated distribution.
Thus, the random walk model produces similar numbers of rotations per sources
as observed in the data and is \emph{consistent with a fraction of sources 
exhibiting no rotations within the typical observing periods of the RoboPol 
program.}

\section{Discussion}
\label{sec:discussion}

Aiming at explaining the polarization variability of a sample of blazars we 
have used a simplistic two-parameter random walk model.
In this framework we assumed that the sources are located in a limited range
of the model parameter space and we have constrained this range through an
optimization process that tries to reproduce a sample of EVPA rotations 
detected in the RoboPol main sample, with rotation parameters similar to the 
observations.
We simulated an entire sample of sources, randomly located in this optimized
parameter space, and tested these simulations against the data.

We have found that our model produces a number of large EVPA rotations
($\Delta\chi \leq 90\,\mathrm{deg}$) comparable to the observations.
The fraction of sources not showing rotations is reproduced by the model as a 
result of the limited duration and number of  periods and a range of
variability time-scales ($0.4$--$38$ cells changing per day in the observer
frame).
This implies that \emph{in the framework of a random walk process we would 
expect to see rotations in all sources, when the observations are long enough.}
Only long-term monitoring could help to further constrain how rarely some 
sources exhibit rotations.

The rotations identified in the RoboPol sample were characterized by four 
parameters -- amplitude, duration, rate, and the variation estimator as a 
measure of smoothness.
The distributions of these parameters produced by the model are not rejected
in more than $74\%$ of the simulations when compared to the observations
using a two-sample KS-test, implying that the model is capable of producing a 
sample of rotations comparable to that from the RoboPol data.
Despite the non-rejecting KS-tests the distributions of some rotation 
parameters are systematically set off from the data.
This offset is more pronounced when comparing just large rotations rather than
rotations with arbitrary amplitudes.
Since the sample of all rotations, which is dominated by small rotations 
($\Delta\chi < 90\,\mathrm{deg}$), is described better by the model than the 
large rotations, this could imply different processes for the smaller 
variability and the periods of larger rotations.
This has been suggested before, e.g. in \citet{2015MNRAS.453.1669B} and 
\citet{2016AA26A...590A..10K}.
Furthermore, it would be consistent with the results presented in 
Blinov~et.~al. (2017, submitted) showing that all large EVPA rotations are 
potentially occurring contemporaneously with gamma-ray flares.
Large rotations may be produced by one process also responsible for gamma-ray
flares, whereas small rotations may be a signature of an underlying stochastic 
process.
The superposition of different processes has also been discussed in 
\citet{2016MNRAS.463.3365A}.
We point out, though, that two processes would not necessarily correspond to a 
strict cut in the rotation amplitude.

The model fails to reproduce distributions of the mean and standard deviation
of the polarization fraction observed in the RoboPol main sample.
The reason is that the model parameter space that is most likely to reproduce
the sample of rotation characteristics is too narrow in the number of cells.
This constrains the mean polarization fraction and its standard deviation.
While we have shown that there is no need to assume two populations of sources
in order to create the apparent dichotomy of rotators and non-rotators reported 
in \citet{2016MNRAS.462.1775B}, two populations centred at different locations 
in the parameter space could broaden the distributions of the polarization 
fraction mean and standard deviation and remove the systematic offsets of the 
simulated rotation parameter ECDFs from the observed ones.

The fact that the entire polarization variability of the RoboPol sample cannot
be consistently explained by our modelling approach implies that
(a) the simplistic random walk model cannot produce this variability in 
general,
(b) it cannot reproduce the variability for all sources, or 
(c) we have not correctly constrained the model parameter space from which we 
randomly draw the simulated source sample.
Therefore, we can certainly not accept the model with the discussed constraints
on the model parameter space.
Neither can we make a strong claim that this random walk model is rejected as 
an explanation for the variability that we observe in the optical polarization.
The fact that this simplistic model and the generic constraints on the 
parameter space can produce several characteristics we have observed implies 
that the hypothesis of a random walk model cannot easily be dismissed and 
should be tested thoroughly.

Complementary to this statistical study that focused on the question whether
the optical polarization variability of the entire RoboPol sample can be 
explained with a simple random walk process, we suggest testing the random
walk hypothesis against each source and rotation event individually.
Following for example the procedure of \citet{2016AA26A...590A..10K} would 
constrain the model parameter space for each source individually without 
requiring an a~priori assumption about the distribution of the number of cells
and the cell variation rate.
Additionally, this approach would allow us to discern whether only particular 
sources or particular rotation events are consistent with the random walk 
model.
Though much more challenging due to the significantly large model parameter
space, more physical models, as presented by \citet{2014ApJ...780...87M}, need
to be tested against the data.
A particular advantage of more sophisticated models is the multi-frequency 
coverage.
Multi-frequency polarization monitoring could put much stronger constraints on
such models.

Our simplistic single-frequency modelling approach demonstrates that the random
walk hypothesis should not be dismissed.
We have pointed out several features in the polarization variability that are 
consistent with the tested random walk process:

\begin{itemize}
  \item Rotations produced by the random walk process have the same 
        dependencies between the characterizing parameters -- amplitude, 
        duration, rate, and variation estimator -- as observed in the data.
  \item During rotation periods the polarization fraction on average decreases.
        In the context of the random walk model rotation periods are not 
        behaving intrinsically differently from other periods.
        By selecting out periods that show an apparent rotation, we select out 
        periods of lower polarization, which are more likely to show apparent 
        rotations.
  \item In the data the amount by which the polarization fraction drops during 
        a rotation correlates with the rotation rate. Though not highly likely,
        it is possible to reproduce this result with the random walk model.
\end{itemize}

\section{Conclusions}
\label{sec:conclusion}

We have searched for rotations of the polarization angle in blazars of the 
RoboPol main sample using three seasons of data.
We have estimated four characteristic parameters for each rotation as well as
the polarization fraction during rotations and showed certain dependencies 
between those parameters.
In a statistical approach, we tested whether a simple random walk model can 
reproduce the entire sample of rotations and their characteristics.
The model fails to reproduce the rotation characteristics and the polarization
fraction at the same time.
But we cannot with certainty state that we have correctly constrained the model 
parameter space.
Therefore, we cannot claim that the random walk model can be rejected.

On the other hand the model succeeds in producing samples of rotations with
characteristic parameters similar to the observed ones.
Furthermore, it is consistent with the apparent dichotomy of rotators and 
non-rotators, the drop of the polarization fraction during rotations, and the 
apparent dependency of the rotation rate on  the polarization fraction 
decrease during rotations.
We have pointed out that testing the model against the data for each source or
each rotation event individually could help to better understand the
limitations of this random walk model and would allow us to test whether only
particular sources or types of rotation events are consistent with a random
walk.
This approach would be complementary to this study that tested whether the 
polarization variability of the entire RoboPol sample could be explained by a
simple random walk model.

The fact that even a very simplified model can reproduce various features that
we observe in the data suggests that the random walk hypothesis should be 
studied in further detail and, in particular, that more physical models should 
be tested on a statistical basis against large data sets such as the RoboPol 
main sample of blazars.


\section*{Acknowledgements}

The RoboPol project is a collaboration between the University of Crete/FORTH 
in Greece, Caltech in the USA, MPIfR in Germany, IUCAA in India and Toru\'{n}
Centre for Astronomy in Poland.
This research was supported in part by NASA grants NNX11A043G and NNX16AR41G 
and NSF grant AST-1109911.
S.K. is supported through NASA grant NNX13AQ89G.
This research was partly funded by the Academy of Finland project 284495.
S.K. particularly thanks T.~Savolainen for comments on this manuscript and 
for the support of this research.
The authors furthermore thank E.~Angelakis, T.~Hovatta, V.~Pavlidou, and 
K.~Tassis from the RoboPol collaboration for comments and intense discussions 
during the collaboration meetings.
This research was done with \verb|python| packages \verb|numpy 1.11.3|,
\verb|scipy 0.18.1|, \verb|statsmodels 0.6.1|, \verb|pytables 3.3.0|,
and \verb|matplotlib 2.0.0|.

\bibliographystyle{mnras}
\bibliography{references}

\begin{thebibliography}{}
\makeatletter
\relax
\def\mn@urlcharsother{\let\do\@makeother \do\$\do\&\do\#\do\^\do\_\do\%\do\~}
\def\mn@doi{\begingroup\mn@urlcharsother \@ifnextchar [ {\mn@doi@}
  {\mn@doi@[]}}
\def\mn@doi@[#1]#2{\def\@tempa{#1}\ifx\@tempa\@empty \href
  {http://dx.doi.org/#2} {doi:#2}\else \href {http://dx.doi.org/#2} {#1}\fi
  \endgroup}
\def\mn@eprint#1#2{\mn@eprint@#1:#2::\@nil}
\def\mn@eprint@arXiv#1{\href {http://arxiv.org/abs/#1} {{\tt arXiv:#1}}}
\def\mn@eprint@dblp#1{\href {http://dblp.uni-trier.de/rec/bibtex/#1.xml}
  {dblp:#1}}
\def\mn@eprint@#1:#2:#3:#4\@nil{\def\@tempa {#1}\def\@tempb {#2}\def\@tempc
  {#3}\ifx \@tempc \@empty \let \@tempc \@tempb \let \@tempb \@tempa \fi \ifx
  \@tempb \@empty \def\@tempb {arXiv}\fi \@ifundefined
  {mn@eprint@\@tempb}{\@tempb:\@tempc}{\expandafter \expandafter \csname
  mn@eprint@\@tempb\endcsname \expandafter{\@tempc}}}

\bibitem[\protect\citeauthoryear{{Aleksi{\'c}} et~al.,}{{Aleksi{\'c}}
  et~al.}{2014a}]{2014A&A...567A..41A}
{Aleksi{\'c}} J.,  et~al., 2014a, \mn@doi [\aap] {10.1051/0004-6361/201323036},
  \href {http://adsabs.harvard.edu/abs/2014A%26A...567A..41A} {567, A41}

\bibitem[\protect\citeauthoryear{{Aleksi{\'c}} et~al.,}{{Aleksi{\'c}}
  et~al.}{2014b}]{2014A&A...569A..46A}
{Aleksi{\'c}} J.,  et~al., 2014b, \mn@doi [\aap] {10.1051/0004-6361/201423484},
  \href {http://adsabs.harvard.edu/abs/2014A%26A...569A..46A} {569, A46}

\bibitem[\protect\citeauthoryear{{Angelakis} et~al.,}{{Angelakis}
  et~al.}{2016}]{2016MNRAS.463.3365A}
{Angelakis} E.,  et~al., 2016, \mn@doi [\mnras] {10.1093/mnras/stw2217}, \href
  {http://adsabs.harvard.edu/abs/2016MNRAS.463.3365A} {463, 3365}

\bibitem[\protect\citeauthoryear{{Blinov} et~al.,}{{Blinov}
  et~al.}{2015}]{2015MNRAS.453.1669B}
{Blinov} D.,  et~al., 2015, \mn@doi [\mnras] {10.1093/mnras/stv1723}, \href
  {http://adsabs.harvard.edu/abs/2015MNRAS.453.1669B} {453, 1669}

\bibitem[\protect\citeauthoryear{{Blinov} et~al.,}{{Blinov}
  et~al.}{2016a}]{2016MNRAS.457.2252B}
{Blinov} D.,  et~al., 2016a, \mn@doi [\mnras] {10.1093/mnras/stw158}, \href
  {http://adsabs.harvard.edu/abs/2016MNRAS.457.2252B} {457, 2252}

\bibitem[\protect\citeauthoryear{{Blinov} et~al.,}{{Blinov}
  et~al.}{2016b}]{2016MNRAS.462.1775B}
{Blinov} D.,  et~al., 2016b, \mn@doi [\mnras] {10.1093/mnras/stw1732}, \href
  {http://adsabs.harvard.edu/abs/2016MNRAS.462.1775B} {462, 1775}

\bibitem[\protect\citeauthoryear{Feigelson \& Babu}{Feigelson \&
  Babu}{2012}]{FeigelsonBabu201208}
Feigelson E.~D.,  Babu G.~J.,  2012, Modern Statistical Methods for Astronomy:
  With R Applications, 1 edn.
Cambridge University Press

\bibitem[\protect\citeauthoryear{{Holmes} et~al.,}{{Holmes}
  et~al.}{1984}]{1984MNRAS.211..497H}
{Holmes} P.~A.,  et~al., 1984, \mn@doi [\mnras] {10.1093/mnras/211.3.497},
  \href {http://adsabs.harvard.edu/abs/1984MNRAS.211..497H} {211, 497}

\bibitem[\protect\citeauthoryear{{Jones}, {Rudnick}, {Aller}, {Aller}, {Hodge}
  \& {Fiedler}}{{Jones} et~al.}{1985}]{1985ApJ...290..627J}
{Jones} T.~W.,  {Rudnick} L.,  {Aller} H.~D.,  {Aller} M.~F.,  {Hodge} P.~E.,
  {Fiedler} R.~L.,  1985, \mn@doi [\apj] {10.1086/163020}, \href
  {http://adsabs.harvard.edu/abs/1985ApJ...290..627J} {290, 627}

\bibitem[\protect\citeauthoryear{{Kiehlmann} et~al.,}{{Kiehlmann}
  et~al.}{2016}]{2016AA26A...590A..10K}
{Kiehlmann} S.,  et~al., 2016, \mn@doi [\aap] {10.1051/0004-6361/201527725},
  \href {http://ui.adsabs.harvard.edu/#abs/2016A%26A...590A..10K} {590, A10}

\bibitem[\protect\citeauthoryear{{Kikuchi}, {Mikami}, {Inoue}, {Tabara}  \&
  {Kato}}{{Kikuchi} et~al.}{1988}]{1988A&A...190L...8K}
{Kikuchi} S.,  {Mikami} Y.,  {Inoue} M.,  {Tabara} H.,   {Kato} T.,  1988,
  \aap, \href {http://adsabs.harvard.edu/abs/1988A%26A...190L...8K} {190, L8}

\bibitem[\protect\citeauthoryear{{King} et~al.,}{{King}
  et~al.}{2014}]{2014MNRAS.442.1706K}
{King} O.~G.,  et~al., 2014, \mn@doi [\mnras] {10.1093/mnras/stu176}, \href
  {http://adsabs.harvard.edu/abs/2014MNRAS.442.1706K} {442, 1706}

\bibitem[\protect\citeauthoryear{{Kinman}}{{Kinman}}{1967}]{1967ApJ...148L..53K}
{Kinman} T.~D.,  1967, \mn@doi [\apjl] {10.1086/180013}, \href
  {http://adsabs.harvard.edu/abs/1967ApJ...148L..53K} {148, L53}

\bibitem[\protect\citeauthoryear{{Konigl} \& {Choudhuri}}{{Konigl} \&
  {Choudhuri}}{1985}]{1985ApJ...289..188K}
{Konigl} A.,  {Choudhuri} A.~R.,  1985, \mn@doi [\apj] {10.1086/162877}, \href
  {http://adsabs.harvard.edu/abs/1985ApJ...289..188K} {289, 188}

\bibitem[\protect\citeauthoryear{{Larionov} et~al.,}{{Larionov}
  et~al.}{2008}]{2008A&A...492..389L}
{Larionov} V.~M.,  et~al., 2008, \mn@doi [\aap] {10.1051/0004-6361:200810937},
  \href {http://adsabs.harvard.edu/abs/2008A%26A...492..389L} {492, 389}

\bibitem[\protect\citeauthoryear{{Larionov} et~al.,}{{Larionov}
  et~al.}{2013}]{2013ApJ...768...40L}
{Larionov} V.~M.,  et~al., 2013, \mn@doi [\apj] {10.1088/0004-637X/768/1/40},
  \href {http://adsabs.harvard.edu/abs/2013ApJ...768...40L} {768, 40}

\bibitem[\protect\citeauthoryear{{Lister} \& {Marscher}}{{Lister} \&
  {Marscher}}{1997}]{1997ApJ...476..572L}
{Lister} M.~L.,  {Marscher} A.~P.,  1997, \apj, \href
  {http://adsabs.harvard.edu/abs/1997ApJ...476..572L} {476, 572}

\bibitem[\protect\citeauthoryear{Longair}{Longair}{2011}]{Longair201103}
Longair M.~S.,  2011, High Energy Astrophysics, 3 edn.
Cambridge University Press, \url {http://amazon.com/o/ASIN/0521756189/}

\bibitem[\protect\citeauthoryear{{Marscher}}{{Marscher}}{2014}]{2014ApJ...780...87M}
{Marscher} A.~P.,  2014, \mn@doi [\apj] {10.1088/0004-637X/780/1/87}, \href
  {http://adsabs.harvard.edu/abs/2014ApJ...780...87M} {780, 87}

\bibitem[\protect\citeauthoryear{{Marscher} et~al.,}{{Marscher}
  et~al.}{2008}]{2008Natur.452..966M}
{Marscher} A.~P.,  et~al., 2008, \mn@doi [\nat] {10.1038/nature06895}, \href
  {http://adsabs.harvard.edu/abs/2008Natur.452..966M} {452, 966}

\bibitem[\protect\citeauthoryear{{Marscher} et~al.,}{{Marscher}
  et~al.}{2010}]{2010ApJ...710L.126M}
{Marscher} A.~P.,  et~al., 2010, \mn@doi [\apjl]
  {10.1088/2041-8205/710/2/L126}, \href
  {http://adsabs.harvard.edu/abs/2010ApJ...710L.126M} {710, L126}

\bibitem[\protect\citeauthoryear{{Nalewajko}}{{Nalewajko}}{2010}]{2010IJMPD..19..701N}
{Nalewajko} K.,  2010, \mn@doi [International Journal of Modern Physics D]
  {10.1142/S0218271810016853}, \href
  {http://adsabs.harvard.edu/abs/2010IJMPD..19..701N} {19, 701}

\bibitem[\protect\citeauthoryear{{Nalewajko} \& {Begelman}}{{Nalewajko} \&
  {Begelman}}{2012}]{2012MNRAS.427.2480N}
{Nalewajko} K.,  {Begelman} M.~C.,  2012, \mn@doi [\mnras]
  {10.1111/j.1365-2966.2012.22117.x}, \href
  {http://adsabs.harvard.edu/abs/2012MNRAS.427.2480N} {427, 2480}

\bibitem[\protect\citeauthoryear{{Nolan} et~al.,}{{Nolan}
  et~al.}{2012}]{2012ApJS..199...31N}
{Nolan} P.~L.,  et~al., 2012, \mn@doi [\apjs] {10.1088/0067-0049/199/2/31},
  \href {http://adsabs.harvard.edu/abs/2012ApJS..199...31N} {199, 31}

\bibitem[\protect\citeauthoryear{{Pavlidou} et~al.,}{{Pavlidou}
  et~al.}{2014}]{2014MNRAS.442.1693P}
{Pavlidou} V.,  et~al., 2014, \mn@doi [\mnras] {10.1093/mnras/stu904}, \href
  {http://adsabs.harvard.edu/abs/2014MNRAS.442.1693P} {442, 1693}

\bibitem[\protect\citeauthoryear{{Storn} \& {Price}}{{Storn} \&
  {Price}}{1997}]{1997Storn}
{Storn} R.,  {Price} K.,  1997, \mn@doi [Journal of Global Optimization]
  {10.1023/A:1008202821328}, 11, 341

\bibitem[\protect\citeauthoryear{{Vlahakis} \& {K{\"o}nigl}}{{Vlahakis} \&
  {K{\"o}nigl}}{2004}]{2004ApJ...605..656V}
{Vlahakis} N.,  {K{\"o}nigl} A.,  2004, \mn@doi [\apj] {10.1086/382670}, \href
  {http://adsabs.harvard.edu/abs/2004ApJ...605..656V} {605, 656}

\bibitem[\protect\citeauthoryear{{Zhang}, {Chen}  \& {B{\"o}ttcher}}{{Zhang}
  et~al.}{2014}]{2014ApJ...789...66Z}
{Zhang} H.,  {Chen} X.,   {B{\"o}ttcher} M.,  2014, \mn@doi [\apj]
  {10.1088/0004-637X/789/1/66}, \href
  {http://adsabs.harvard.edu/abs/2014ApJ...789...66Z} {789, 66}

\bibitem[\protect\citeauthoryear{{Zhang}, {Chen}, {B{\"o}ttcher}, {Guo}  \&
  {Li}}{{Zhang} et~al.}{2015}]{2015ApJ...804...58Z}
{Zhang} H.,  {Chen} X.,  {B{\"o}ttcher} M.,  {Guo} F.,   {Li} H.,  2015,
  \mn@doi [\apj] {10.1088/0004-637X/804/1/58}, \href
  {http://adsabs.harvard.edu/abs/2015ApJ...804...58Z} {804, 58}

\makeatother
\end{thebibliography}


\bsp	
\label{lastpage}
\end{document}